\documentclass[aps,prb,eqsecnum]{revtex4}

\newcommand{\diag}{{\rm diag\,}}

\newcommand{\Tr}{{\rm Tr\,}}

\newcommand{\detg}{{\rm detg\,}}

\newcommand{\ul}[1]{\underline{#1}}
\newcommand{\xu}[2]{x_{#1}^{(#2)}}
\newcommand{\su}[2]{s_{#1 1}^{(#2)}}
\newcommand{\ee}{{(1)}}

\begin{document}

\title{Angular Gelfand--Tzetlin Coordinates
       for the Supergroup ${\rm UOSp}(k_1/2k_2)$}

\author{Thomas Guhr$^1$ and Heiner Kohler$^2$}

\affiliation{$^1$ Matematisk Fysik, LTH, Lunds Universitet,
                  Box 118, 22100 Lund, Sweden\\
             $^2$ Departamento de Teor\'{i}a de Materia
                   Condensada, Universidad
                  Aut\'onoma, Madrid, Spain}

\date{\today}

\begin{abstract}
  We construct Gelfand--Tzetlin coordinates for the unitary
  orthosymplectic supergroup ${\rm UOSp}(k_1/2k_2)$. This extends a
  previous construction for the unitary supergroup ${\rm U}(k_1/k_2)$.
  We focus on the angular Gelfand--Tzetlin coordinates, i.e.~our
  coordinates stay in the space of the supergroup.  We also present a
  generalized Gelfand pattern for the supergroup ${\rm
  UOSp}(k_1/2k_2)$ and discuss various implications for representation
  theory.
\end{abstract}

\maketitle

\section{Introduction}
\label{sec1}

If the symmetries of a physical problem are simple enough, proper
coordinates are easy to find. However, already the Schr\"odinger
equation for a particle in a potential with spherical symmetry leads
to non--trivial group theory, such as parametrization of the Lie group
${\rm SO}(3)$ with Euler angles, spherical harmonics, Wigner
representation functions and, in the case of the Hydrogen atom,
additional symmetries and the Lie group ${\rm SO}(4)$. The coordinates
mostly used distinguish, for a good physics reason, certain directions
and thus do not treat all coordinates on an equal footing. Group
theoretically, such parametrizations are called non--canonical. The
Euler angles, for example, describe three subsequent rotations, first,
about the $z$--axis, second, about the new $y$--axis, and, third,
about the new $z$--axis.  Nevertheless, there are many problems,
particularly in statistical mechanics, in many--body physics and in
matrix models, where one does not want to distinguish certain
directions. Rather, all variables parametrizing the group should be
treated on an equal footing.  Gelfand--Tzetlin
coordinates~\cite{GT,GT2,BR} are such a coordinate system. Their
construction is based on a group chain or coset decomposition. Thus
Gelfand--Tzetlin coordinates have a clear recursive structure.  From a
physics point of view, it is important that matrix elements, measures
and other quantities reflect this clear recursive structure and can be
given very explicitly.  The generality of this group chain
construction makes Gelfand--Tzetlin coordinates powerful tools in
applications, see for example Refs.~\onlinecite{SLS,GGT}, but also for
conceptual studies, see for example Refs.~\onlinecite{GS1,GS2,AFS}. A
particularly intriguing aspect is the intimate and direct connection
between Gelfand--Tzetlin coordinates on the group manifold and
representations of this group. Their rich features and their relevance
for different types of studies ranging from physics applications to
pure mathematics render Gelfand--Tzetlin coordinates important objects
in their own right.

In Ref.~\onlinecite{GGT}, Gelfand--Tzetlin coordinates were
constructed for the unitary supergroup ${\rm U}(k_1/k_2)$. In the
present contribution, we further extend this and construct
Gelfand--Tzetlin coordinates for the unitary orthosymplectic
supergroup ${\rm UOSp}(k_1/2k_2)$.  As this group contains the
symplectic group ${\rm USp}(2k_2)$ and the orthogonal group ${\rm
SO}(k_1)$ as subgroups, our construction also includes coordinate
systems for these two groups in ordinary space. The construction for
the orthogonal group was implicitly also done in
Ref.~\onlinecite{AFS}.

For the sake of clarity, an important remark is in order: We
distinguish between {\it angular} and {\it radial} Gelfand--Tzetlin
coordinates. In the present work, we construct angular ones.  By that
we mean, that they never leave the space of the group and its
algebra. In previous contributions~\cite{GUKOP1,GUKO1,GUKOP2,GUKO2},
we constructed radial Gelfand--Tzetlin coordinates to study certain
types of group integrals. These radial Gelfand--Tzetlin coordinates
are capable of mapping the integral over a group onto integrals over
the radial part of a different symmetric space. Hence, in this sense,
these coordinates leave the space of the group and its algebra.  Here,
we always stay with the angular Gelfand--Tzetlin coordinates.

The appreciated explicit formulae resulting from the Gelfand--Tzetlin
construction imply the unavoidable disadvantage that a reader, not
familiar with the subject, can quickly lose his
orientation. Therefore, we decided to skip several detailed
calculations if, in our opinion, it would not be too cumbersome for
the reader to recover the missing steps by properly adjusting the
corresponding ones in Ref.~\onlinecite{GGT}. In any case, we recommend
that an interested but unexperienced reader studies first
Refs.~\onlinecite{SLS,AFS} and then Ref.~\onlinecite{GGT} before
reading the present contribution.

The paper is organized as follows: In Sec.~\ref{sec2}, we construct
the angular Gelfand--Tzetlin coordinates.  We state the generalized
Gelfand pattern in Sec.~\ref{sec3} and discuss some issues related to
representation theory.  Summary and conclusions are given in
Sec.~\ref{sec4}.

\section{Construction of the coordinate system}
\label{sec2}

In Sec.~\ref{sec21}, we collect some properties of the supergroup
${\rm UOSp}(k_1/2k_2)$ needed in the sequel.  We set up the proper
Gelfand--Tzetlin equations and their recursion to all levels in
Secs.~\ref{sec22} and~\ref{sec23}, respectively.  We solve these
equations in Sec.~\ref{sec24}.  We summarize the construction of the
Gelfand--Tzetlin coordinates for the ordinary unitary symplectic group
in Sec.~\ref{sec25}.  The invariant measure of the supergroup is
worked out in Sec.~\ref{sec26}. The matrix elements of the supergroup
are obtained in Sec.~\ref{sec27}.

\subsection{The supergroup ${\rm UOSp}(k_1/2k_2)$}
\label{sec21}

The classification of superalgebras and supergroups can be found in
Refs.~\onlinecite{KAC1,KAC2,RIT}. Here, we restrict ourselves to
summarizing features of the supergroups ${\rm OSp}(k_1/2k_2)$ and
${\rm UOSp}(k_1/2k_2)$. We will refer to $k_1$ and $2k_2$ as to the
bosonic and fermionic dimensions, respectively. We introduce the
notation $(k_1/2k_2)$ for the resulting superdimension. The elements
of ${\rm OSp}(k_1/2k_2)$ are those elements $u$ of the general linear
supergroup ${\rm GL}(k_1/2k_2)$ which satisfy $u^\dagger Lu=L$. The
metric $L$ is given by
\begin{equation}
L = \diag(1_{k_1},1_{k_2}\otimes\tau^{(1)}) \ ,
\label{metL}
\end{equation}
where $1_{k_1}$ and $1_{k_2}$ are the $k_1\times k_1$ and the
$k_2\times k_2$ unit matrices and where $\tau^{(1)}$ is one of the
Pauli matrices,
\begin{eqnarray}
\tau^{(0)} \ = \ \left[ \begin{array}{cc}
                        1 & 0 \\
                        0 & 1
                        \end{array} \right] \ , \qquad
\tau^{(1)} \ = \ \left[ \begin{array}{cc}
                         0 & +1 \\
                        -1 &  0
                        \end{array} \right] \ , \nonumber\\
\tau^{(2)} \ = \ \left[ \begin{array}{cc}
                         0 & -i \\
                        -i &  0
                        \end{array} \right] \ , \qquad
\tau^{(3)} \ = \ \left[ \begin{array}{cc}
                        +i &  0 \\
                         0 & -i
                        \end{array} \right] \ .
\label{Pauli}
\end{eqnarray}
The supergroup ${\rm UOSp}(k_1/2k_2)$ is the compact subgroup of ${\rm
OSp}(k_1/2k_2)$. By construction, the direct product ${\rm
SO}(k_1)\otimes {\rm USp}(2k_2)$ of the ordinary orthogonal and the
ordinary unitary symplectic group is a subgroup of ${\rm
UOSp}(k_1/2k_2)$. As is well known, the ordinary orthogonal group
${\rm SO}(k_1)$ has slightly different features for even and odd
dimension $k_1$. Thus, these differences are also present in the
supergroup ${\rm UOSp}(k_1/2k_2)$.

The group elements act on a graded space, which we denote by ${\cal
L}=^0{\cal L}\oplus^1{\cal L}$. It decomposes into a sum of an even
$^0{\cal L}$ and an odd $^1{\cal L}$ subspace according to its
transformation properties under the parity automorphism~\cite{BER}. We
define a basis $e_j= e_{j1}, j=1, \ldots k_1$ for $^0{\cal L}$, and
$e_{k_1+j}= e_{j2}, j=1, \ldots2k_2$ for $^1{\cal L}$ respectively.

The supergroup ${\rm UOSp}(k_1/2k_2)$ can be obtained by the
exponential mapping of the superalgebra ${\rm uosp}(k_1/2k_2)$, such
that $\sigma\in {\rm uosp}(k_1/2k_2)$ leads to $u=\exp(\sigma)\in {\rm
UOSp}(k_1/2k_2)$. The construction of the angular Gelfand--Tzetlin
coordinates uses as the starting point the Cartan subalgebra ${\rm
uosp}^{(0)}(k_1/2k_2)$ of ${\rm uosp}(k_1/2k_2)$. For even bosonic
dimension $2k_1$, the elements of ${\rm uosp}^{(0)}(2k_1/2k_2)$ are
the matrices
\begin{equation}
s = \diag(is_{11}\tau^{(1)},\ldots,is_{k_11}\tau^{(1)},
          s_{12}\tau^{(3)},\ldots,s_{k_21}\tau^{(3)}) \ ,
\label{csue}
\end{equation}
while for odd bosonic dimension $2k_1+1$,
the ${\rm uosp}^{(0)}(2k_1+1/2k_2)$
consists of the matrices
\begin{equation}
s = \diag(is_{11}\tau^{(1)},\ldots,is_{k_11}\tau^{(1)},0,
    s_{12}\tau^{(3)},\ldots,s_{k_21}\tau^{(3)}) \ . \label{csuo}
\end{equation}
Naturally, ${\rm uosp}^{(0)}(k_1/2k_2)$ is the direct sum of the Cartan
subalgebras of ${\rm so}(k_1)$ and ${\rm usp}(2k_2)$.

\subsection{Derivation of the angular Gelfand--Tzetlin equations}
\label{sec22}

Gelfand--Tzetlin coordinates are based on a group chain or,
equivalently, on a coset decomposition. The coset decomposition needed
for the supergroup ${\rm UOSp}(k_1/2k_2)$ is
\begin{eqnarray}
{\rm UOSp}(k_1/2k_2) &=&
    \frac{{\rm UOSp}(k_1/2k_2)}
         {{\rm UOSp}\left((k_1-1)/2k_2\right))}
    \otimes \frac{{\rm UOSp}\left((k_1-1)/2k_2\right)}
                 {{\rm UOSp}\left((k_1-2)/2k_2)\right)}
    \otimes \cdots \otimes \frac{{\rm UOSp}(1/2k_2)}
                                {{\rm USp}(2k_2)} \nonumber\\
         & & \qquad \otimes \frac{{\rm USp}(2k_2)}
                                 {{\rm USp}(2k_2-2)}
             \otimes \cdots \otimes \frac{{\rm USp}(4)}
                                         {{\rm SU}(2)}
                    \otimes {\rm SU}(2) \quad .
\label{gt0.1}
\end{eqnarray}
Every coset space describes a unit sphere. The first coset ${\rm
UOSp}(k_1/2k_2)/{\rm UOSp}\left((k_1-1)/2k_2\right))$ is a sphere in a
superspace with dimension $(k_1/2k_2)$. The dimension of the space in
which the sphere lives is lowered by one in every step. The sphere
${\rm UOSp}(1/2k_2)/{\rm USp}(2k_2)$ is the last one living in a
superspace, the following spheres in the second line of
Eq.~(\ref{gt0.1}) are spheres in ordinary spaces.  Coordinate systems
will be constructed on all these spheres under the non--trivial
requirement that the orthogonality, more precisely the equation
$u^\dagger Lu=L$, is always respected.  Thus, once the coordinate
system on one sphere has been obtained, the orthogonal complement to
every fixed vector on this sphere has to be constructed, the next
sphere lives in this smaller space. Hence, loosely speaking, the
spheres in the coset decomposition are orthogonal to each other.  The
construction to follow is an extension of the one in
Ref.~\onlinecite{GGT} for the unitary supergroup.  For simplicity, we
consider the case of even $k_1$ first. The differences occurring for
odd $k_1$ will be dealt with in Sec.~\ref{sec23}.

To project onto a smaller subspace, we write $u\in {\rm UOSp}(k_1/2
k_2)$ as $u=[u_1 \, u_2 \cdots u_{k_1+2 k_2}]$ where the columns $u_i$
are normalized supervectors. We denote by $u_{ji}$ their entries in
the basis $e_{j1}, j=1, \ldots k_1$ and $ e_{j2}, j=1,
\ldots2k_2$. The orthogonality condition requires the vectors $u_i$,
$i\leq k_1$ to be real
\begin{equation}
u_{ji}=u_{ji}^* \quad {\rm ,\quad for}\quad 1\leq j\leq k_1 \quad {\rm and}\quad
u_{(k_1+2j)i}=u_{(k_1+2j-1)i}^* \quad {\rm ,\quad for\quad } 1\leq j\leq k_2.
\label{gt0}
\end{equation}
We consider the first vector, it is parametrized by $k_1$ real
commuting variables $u_{j1}$ and $2k_2$ complex anticommuting
variables, for the latter we write
\begin{equation}
u_{(k_1+2j)i}=\alpha_j^{\ast} \ , \quad 1\leq j\leq k_2 \ . \label{gt0a}
\end{equation}
We also define $|\alpha_j|^2=\alpha_j^\ast\alpha_j$.  The supervector
$u_1$ describes the coset space ${\rm UOSp}(k_1/2k_2)/{\rm
UOSp}((k_1-1)/2k_2)$ which is -- similar to ordinary spaces --
isomorphic to the surface of the $(k_1/2k_2)$ dimensional sphere
$S^{(k_1-1)/2k_2}$.  We go from Cartesean coordinates to a new set of
coordinates for $u_1$ by projecting a fixed element $s$ of the Cartan
subalgebra on a space of superdimension $((k_1-1)/2 k_2)$ orthogonal
to $u_1$,
\begin{equation}
s^{(1)} = (1_{k_1+2k_2}-u_1u_1^\dagger)s(1_{k_1+2k_2}-u_1u_1^\dagger) \ .
\label{gt0b}
\end{equation}
The eigenvalues and eigenvectors of this projected matrix are obtained
by solving the supersymmetric Gelfand--Tzetlin equation
\begin{equation}
s_p^{(1)} e_{p}^{(1)} =
(1_{k_1+2k_2}-u_1u_1^\dagger)s(1_{k_1+2k_2}-u_1u_1^\dagger)e_{p}^{(1)} =
(1_{k_1+2k_2}-u_1u_1^\dagger)se_{p}^{(1)} \ ,
\label{gt1}
\end{equation}
which extends the equation in Ref.~\onlinecite{GGT} for the unitary
supergroup to ${\rm UOSp}(k_1/2k_2)$. It is convenient to rotate the
basis in such a way that $s$ becomes diagonal before solving
Eq.~(\ref{gt1}), we introduce the primed basis
\begin{eqnarray}
e_{(2i-1)1}^{\prime}&=&\frac{1}{\sqrt{2}}\left(e_{(2i-1)1}+
                       i e_{(2i)1}\right) \ , \nonumber\\
e_{(2i)1}^{\prime}&=&\frac{1}{\sqrt{2}}
                 \left(ie_{(2i-1)1}+e_{(2i)1}\right) \ ,
\quad i=1\ldots k_1/2 \ , \nonumber\\
e_{i2}^{\prime}&=&e_{i2} \ , \quad i=k_1+1, \ldots k_1+2k_2 \ .
\label{gt2}
\end{eqnarray}
The rotation only affects the bosonic degrees of freedom, not the
fermionic ones.  Due to this rotation, the bosonic entries of
$u_1^\prime$ are now complex variables which we write in the form
\begin{equation}
u_{(2j)1}^\prime=iu_{(2j-1)1}^{\prime\ast}=
\frac{i}{\sqrt{2}}|v_j^{(1)}|
     \exp\left(-i\vartheta_j^{(1)}\right) \ , \quad
                  j=1,\ldots,k_1/2\ .
\label{gt1.1}
\end{equation}
The fermionic entries are, also in the primed basis, given by
Eq.~(\ref{gt0a}). To calculate the eigenvalues, we need the
characteristic function of the eigenvalue equation (\ref{gt1}),
\begin{eqnarray}
z\left(s_p^{(1)}\right) =
   \detg\left(\left(1_{k_1+2k_2}-u_1 u_1^\dagger\right)s
                      -s_p^{(1)}\right)
 =  -s_p^{(1)}  \detg\left(s-s_p^{(1)}\right)
                   u_1^\dagger\frac{1_{k_1+2k_2}}{s-s_p^\ee}u_1 \ .
\label{gt3}
\end{eqnarray}
Importantly, the function $z(s_p^{(1)})$ behaves differently for the
$k_1$ bosonic eigenvalues, i.e.~for those in the boson--boson block
$s_p^{(1)}=s_{p1}^{(1)}, \ p=1,\ldots,k_1$, and for the $2k_2$
fermionic eigenvalues, i.e.~for those in the fermion--fermion block
$s_{k_1+p}^{(1)}=is_{p2}^{(1)}, \ p=1,\ldots,2k_2$. The equation above
has therefore to be discussed in the limits
\begin{eqnarray}
z\left(s_p^\ee\right) \ \longrightarrow \
          \cases{ 0       & \ for \  $p=1,\ldots,k_1$ \cr
                  \infty  & \ for \  $p=k_1+1,\ldots,\ k_1+2k_2$ } \quad .
\label{gt4}
\end{eqnarray}
Together with the normalization condition $u_1^\dagger u_1=1$ we
find the following set of equations,
\begin{eqnarray}
1 & = & \sum_{p=1}^{k_1/2} |v_p^\ee|^2
         +  \sum_{p=1}^{k_2} |\alpha_p^\ee|^2  , \label{gt50}\\
0 & = & (s_{p1}^{(1)})^2
        \left(\sum_{q=1}^{k_1/2}
        \frac{|v_q^{(1)}|^2}{(s_{q1})^2-(s_{p1}^{(1)})^2} +
        \sum_{q=1}^{k_2}
        \frac{|\alpha_q^\ee|^2}
             {(is_{q2})^2-(s_{p1}^{(1)})^2}
        \right) \ , \qquad
        p=1,\ldots,(k_1-1) \ , \label{gt51}\\
z_p & = & (is_{p2}^{(1)})^2
          \frac{\prod_{q=1}^{k_1/2}
          \left((s_{q1})^2-(is_{p2}^{(1)})^2\right)}
          {\prod_{q=1}^{k_2}
          \left((is_{q2})^2-(is_{p2}^{(1)})^2\right)}
          \left(\sum_{q=1}^{k_1/2}
          \frac{|v_q^\ee|^2}{(s_{q1})^2-(is_{p2}^{(1)})^2} +
          \sum_{q=1}^{k_2}
         \frac{|\alpha_q^\ee|^2}{(is_{q2})^2-(is_{p2}^{(1)})^2}
         \right) \ , \nonumber\\
    &  & \qquad\qquad\qquad\qquad\qquad\qquad\qquad\qquad\qquad
                         z_p\to\infty, \quad p=1,\ldots,2k_2 \ .
\label{gt52}
\end{eqnarray}
This is a system of equations in the variables $(s_{p1}^{(1)})^2$ and
$(is_{p2}^{(1)})^2$. The second equation has a twofold degenerate
solution at $s_{p1}^{(1)}=0$. If $s_{p1}^{(1)},is_{p2}^{(1)}$ are
solutions of the above equations, $-s_{p1}^{(1)}$ and $-is_{p2}^{(1)}$
are solutions as well. Hence the projected matrix~(\ref{gt0b}) is of
the form~(\ref{csuo}) in the proper basis $e_j^{(1)}$ and belongs
itself to the Cartan subalgebra ${\rm uosp}^{(0)}((k_1-1)/2k_2)$. This
is crucial for the recursion. The system (\ref{gt50}) to (\ref{gt52})
is overdetermined, out of the $k_1+2k_2+1$ equations in (\ref{gt50})
to (\ref{gt52}), only $k_1/2+k_2$ are independent. The system yields
the moduli squared of the entries of the vector $u_1^{\prime}$
expressed in terms of the eigenvalues $s^{(1)}$. We call the latter
bosonic eigenvalues, if they satisfy Eq.~(\ref{gt51}) and fermionic
eigenvalues if they satisfy Eq.~(\ref{gt52}).  With the substitutions
$s_{q1}^{(j)}\rightarrow(s_{q1}^{(j)})^2$ and
$is_{q2}^{(j)}\rightarrow(is_{q2}^{(j)})^2$, $j=1,2$ , the set of
independent equations is equivalent to the corresponding set of
equations for the unitary supergroup.  Thus, we can directly read off
the solutions from Ref.~\onlinecite{GGT}. They will be stated in
Sec.~\ref{sec24}.

\subsection{Recursion to all levels in superspace}
\label{sec23}

The construction just outlined for the first coset space has to be
continued recursively to cover the entire group manifold.  For the
ordinary groups, this recursion can be found in
Ref.~\onlinecite{GT,GT2,BR}. In the present case, we extend the
recursion for the unitary supergroup in Ref.~\onlinecite{GGT}. As the
Cartan subalgebra ${\rm uosp}^{(0)}(k_1/2k_2)$ is slightly different
for even and odd bosonic dimension $k_1$ according to
Eqs.~(\ref{csue}) and~(\ref{csuo}), we have to distinguish these two
cases for the recursion. For brevity, we refer to a level as even, if
$(k_1-n+1)$ is even, and as odd otherwise.

In the $n$--th step the vector $u_n^\prime$ is expanded in a set of
$k_1-n+1+2k_2$ basis vectors $\mbox{$e_j^\prime$}^{(n-1)}$, which span
the subspace of ${\cal L}$ orthogonal to $u=[u_1 \, u_2 \cdots
u_{n-1}]$. This set splits into two disjoint subsets. The first subset
contains $k_1-n+1$ vectors $\mbox{$e_{j1}^\prime$}^{(n-1)}$ spanning
some subspace of $^0{\cal L}$.  The second one contains $2k_2$ basis
vectors $\mbox{$e_{j2}^\prime$}^{(n-1)}$ spanning $^1{\cal L}$. The
entries of $u_n$ in this basis are complex variables
\begin{eqnarray}
\mbox{$e_{(2p)1}^\prime$}^{(n-1)\dagger} u_n
 &=& i\left(\mbox{$e_{(2p-1)1}^\prime$}^{(n-1)\dagger} u_n\right)^\ast
 = \frac{i}{\sqrt{2}} |v^{(n)}_p|\exp\left(-i\vartheta_p^{(n)}\right) \ ,
 \quad p \leq \cases{(k_1-n+1)/2 & \ for \ $k_1-n+1$ \ even \cr
                     (k_1-n)/2 & \ for \  $k_1-n+1$ \ odd} \ ,
                              \nonumber\\
\mbox{$e_{(2p)2}^\prime$}^{(n-1)\dagger} u_n
 &=& \left(\mbox{$e_{(2p-1)1}^\prime$}^{(n-1)\dagger} u_n\right)^\ast
= \frac{1}{\sqrt{2}} \alpha_p^{(n)\ast} \ , \quad p \leq k_2 \ .
\label{rec2}
\end{eqnarray}
For $k_1-n+1$ odd, the remaining entry is parametrized by a real
variable and an integer $r\in \{ 0,1 \}$ as
\begin{equation}
\mbox{$e_{(k_1-n+1)1}^\prime$}^{(n-1)\dagger} u_n
= (-1)^r v_{\frac{k_1-n}{2}+1}^{(n)}\quad .
\label{rec3}
\end{equation}
Thus, we can write down the rotated $n$--th eigenvector on the
$(n-1)$--th level
\begin{equation}
u_n^{\prime(n-1)}= \sum_{j=1}^{k_1+2k_2-n+1}e_j^{\prime (n-1)\dagger}u_n
                   e_j^{\prime (n-1)}\ .
\label{rec4}
\end{equation}
The projection of $s$ onto this subspace after the $n$--th step is
given by
\begin{equation}
s^{(n-1)}=
\left(\sum_{i=n}^{k_1+2k_2}u_i u_i^\dagger\right) s
\left(\sum_{i=n}^{k_1+2k_2}u_i u_i^\dagger\right) =
\left(1_{k_1+2k_2}-\sum_{i=1}^{n-1}u_i u_i^\dagger\right) s
\left(1_{k_1+2k_2}-\sum_{i=1}^{n-1}u_i u_i^\dagger\right)
\label{rec5}
\end{equation}
and belongs to the Cartan subalgebra of ${\rm UOSp}\left((k_1-n)/2
k_2\right)$. The new coordinates are obtained by projecting
$s^{(n-1)}$ on the subspace orthogonal to $u_n$ by
\begin{equation}
s_p^{(n)}  e_p^{(n)} =
(1_{k_1+2k_2}-u_n u_n^\dagger)s^{(n-1)}(1_{k_1+2k_2}-u_n u_n^\dagger)
             e_p^{(n)} =
(1_{k_1+2k_2}-u_n u_n^\dagger)s^{(n-1)} e_p^{(n)} \ .
\label{rec6}
\end{equation}
For $k_1-n+1$ even, this leads to a system of equations as in
(\ref{gt50}) to~(\ref{gt52}) reduced by $(n-1)/2$ unknown variables.
For $(k_1-n+1)$ odd, the equations have a slightly different form,
\begin{eqnarray}
1 &=& \sum_{p=1}^{\frac{k_1-n}{2}} |v_p^{(n)}|^2
+|v_{\frac{k_1-n}{2}+1}^{(n)}|^2
+  \sum_{p=1}^{k_2} |\alpha_p^{(n)}|^2 \ , \label{rec7a}\\
0 &=& s_{p1}^{(n)} \left(\sum_{q=1}^{\frac{k_1-n}{2}}
    \frac{(s_{p1}^{(n)})^2
      |v_q^{(n)}|^2}{(s_{q1}^{(n-1)})^2-(s_{p1}^{(n)})^2}  +
    |v_{\frac{k_1-n}{2}+1}^{(n)}|^2  +
    \sum_{q=1}^{k_2} \frac{(s_{p1}^{(n)})^2|\alpha_q^{(n)}|^2}
       {(is_{q2}^{(n-1)})^2-(s_{p1}^{(n)})^2}\right) \ , \quad
  p=1,\ldots,(k_1-n) \ , \label{rec7b}\\
z_p &=& -is_{p2}^{(n)}\frac{\prod_{q=1}^{\frac{k_1-n}{2}}
\left((s_{q1}^{(n-1)})^2-(is_{p2}^{(n)})^2\right)}
{\prod_{q=1}^{k_2}\left((is_{q2}^{(n-1)})^2-(is_{p2}^{(n)})^2\right)}
 \left(\sum_{q=1}^{\frac{k_1-n}{2}}
  \frac{(s_{p1}^{(n)})^2|v_q^{(n)}|^2}
{(s_{q1}^{(n-1)})^2-(is_{p2}^{(n)})^2}  +
  |v_{\frac{k_1-n}{2}+1}|^2  +
     \sum_{q=1}^{k_2} \frac{(s_{p1}^{(n)})^2|\alpha_q^{(n)}|^2}
{(is_{q2}^{(n-1)})^2-(is_{p2}^{(n)})^2}\right) \ , \nonumber\\
    & & \qquad\qquad\qquad\qquad\qquad\qquad\qquad\qquad\qquad
                         z_p\to\infty, \quad p=1,\ldots,2k_2 \ .
\label{rec7}
\end{eqnarray}
The difference between Eqs.~(\ref{rec7a}) to~(\ref{rec7}) and the
corresponding equations (\ref{gt50}) to (\ref{gt52}) for the even
levels is due to the isolated entry (\ref{rec3}), which has to be
treated separately. This reflects the difference between the even
orthogonal group and the odd orthogonal group in ordinary space.

The new basis vectors $\mbox{$e_{j}^\prime$}^{(n)}$ are related to the
basis vectors of the foregoing level by a
$(k_1-n+2k_2)\times(k_1-n+1+2k_2)$ rectangular supermatrix
$\mbox{$\widehat{b}^{\prime}$}^{(n)}$. The moduli squared of its
entries $\mbox{$\widehat{b}_{pm}^{\prime}$}^{(n)}$ are determined by
rewriting Eq.~(\ref{rec6}) and multiplying it from the left hand side
with $\mbox{$e_m^\prime$}^{(n-1)\dagger}$
\begin{equation}
\mbox{$\mbox{$e_m^\prime$}^{(n-1)}$}^\dagger s^{(n-1)}
\mbox{$e_p^\prime$}^{(n)}=
s_p^{(n)}\mbox{$\mbox{$e_m^\prime$}^{(n-1)}$}^\dagger
\mbox{$e_p^\prime$}^{(n)}+
\mbox{$\mbox{$e_m^\prime$}^{(n-1)}$}^\dagger u_nb_p^{(n)} \qquad ,
\label{rec8}
\end{equation}
where we defined $ b_p^{(n)}=
u_n^{\dagger}s^{(n-1)}\mbox{$e_p^\prime$}^{(n)}$.  On the other hand
we have
\begin{equation}
\mbox{$\mbox{$e_m^\prime$}^{(n-1)}$}^\dagger s^{(n-1)}
\mbox{$e_p^\prime$}^{(n)}=
s_m^{(n-1)}\mbox{$\mbox{$e_m^\prime$}^{(n-1)}$}
\mbox{$e_p^\prime$}^{(n)}\quad ,
\label{rec9}
\end{equation}
which yields for the matrix elements of
$\mbox{$\widehat{b}^{\prime}$}^{(n)}$ the expression
\begin{equation}
\mbox{$\widehat{b}_{pm}^\prime$}^{(n)}=\frac{1}{s_m^{(n-1)}-s_p^{(n)}}
             \mbox{$u^\prime_{mn}$}^{(n-1)}b_p^{(n)}\quad .
\label{rec10}
\end{equation}
The modulus squared of $b_p^{(n)}$ is determined by the normalization
of the rotated basis vectors $\mbox{$e_m^\prime$}^{(n)\dagger}
\mbox{$e_p^\prime$}^{(n)}= \delta_{mp}$, i.e.~by the condition that
the matrix
$\mbox{$\widehat{b}^{\prime}$}^{(n)}\mbox{$\widehat{b}^{\prime}$}^{(n)\dagger}$
is unity in the $k_1-n+2k_2$ dimensional subspace orthogonal to
$u=[u_1 \, u_2 \cdots u_n]$. Due to the block structure of the
supermatrix $\mbox{$\widehat{b}^{\prime}$}^{(n)}$, the vector
$b^{(n)}$ has commuting and anticommuting elements. For $k_1-n+1$ even
we define $|w_p^{(n)}|^2=|b_{2p}^{(n)}|^2=|b_{2p-1}^{(n)}|^2 \ ,
p=1,\ldots,(k_1-n+1)/2$ for the commuting and
$|\beta_p^{(n)}|^2=|b_{k_1-n+1+2p}^{(n)}|^2=|b_{k_1-n+2p}^{(n)}|^2,
p=1,\ldots,k_2$ for the anticommuting elements. For $k_1-n+1$ odd we
define $|w_p^{(n)}|^2$ and $|\beta_p^{(n)}|^2$ correspondingly.
Again, there is a difference in the determining equations of
$|w_p^{(n)}|^2$ and $|\beta_p^{(n)}|^2$ between the even and the odd
levels of the recursion. For $k_1-n+1$ even we have
\begin{eqnarray}
\frac{1}{|w_{p}^{(n)}|^2}
           &= &\sum_{m=1}^{(k_1-n+1)/2}
              \frac{(s_{m1}^{(n-1)})^2+(s_{p1}^{(n)})^2}
              {\left((s_{m1}^{(n-1)})^2-(s_{p1}^{(n)})^2\right)^2}
              |v_m^{(n)}|^2+
              \sum_{m^\prime=1}^{k_2}
              \frac{(is_{m^\prime2}^{(n-1)})^2+(s_{p1}^{(n)})^2}
       {\left((is_{m^\prime2}^{(n-1)})^2-(s_{p1}^{(n)})^2\right)^2}
             |\alpha_{m^\prime}^{(n)}|^2 , \nonumber\\
          & &\hspace{6cm} p=1,\ldots,\frac{k_1-n-1}{2} \ .
\label{rec11}
\end{eqnarray}
For the remaining modulus squared we obtain
\begin{equation}
\frac{1}{|w_{\frac{k_1-n+1}{2}}^{(n)}|^2} =
                    \sum_{m=1}^{(k_1-n+1)/2}
                    \frac{1}{(s_{m1}^{(n-1)})^2}|v_m^{(n)}|^2+
                    \sum_{m^\prime=1}^{k_2}
                    \frac{1}{(is_{m^\prime2}^{(n-1)})^2}
                    |\alpha_{m^\prime}^{(n)}|^2 \qquad.
\label{rec12}
\end{equation}
The moduli squared of the anticommuting coordinates of $b^{(n)}$
fulfil a formally similar equation. However, it is mathematically more
precise to write it in the inverted form to avoid the appearance of
purely nilpotent variables in the denominator,
\begin{eqnarray}
1 &=& |\beta_{p}^{(n)}|^2
          \left(\sum_{m=1}^{(k_1-n+1)/2}
              \frac{(s_{m1}^{(n-1)})^2+(is_{p2}^{(n)})^2}
             {\left((s_{m1}^{(n-1)})^2-(is_{p2}^{(n)})^2\right)^2}
              |v_m^{(n)}|^2+
              \sum_{m^\prime=1}^{k_2}
              \frac{(is_{m^\prime2}^{(n-1)})^2+(s_{p1}^{(n)})^2}
        {\left((is_{m^\prime2}^{(n-1)})^2-(s_{p1}^{(n)})^2\right)^2}
             |\alpha_{m^\prime}^{(n)}|^2\right)\ , \nonumber\\
    & & \qquad\qquad\qquad\qquad\qquad\qquad\qquad\qquad\qquad
                             p=1,\ldots,k_2 \ .
\label{rec13}
\end{eqnarray}
The corresponding equations for the odd levels are obtained from
Eqs.~(\ref{rec11}) and (\ref{rec13}) by making the following formal
replacements. In Eq.~(\ref{rec11}), the sum over $m$ runs only to
$(k_1-n)/2$ and, in addition, the term
$|v_{\frac{k_1-n}{2}+1}^{(n)}|^2/(s_{p1}^{(n)})^2$ is subtracted. In
Eq.~(\ref{rec13}), the first sum runs only to $(k_1-n)/2$ and the term
$|v_{\frac{k_1-n}{2}+1}^{(n)}|^2/(is_{p2}^{(n)})^2$ is
subtracted. Moreover, Eq.~(\ref{rec12}) does not exist for the odd
levels.

\subsection{Solution of the angular Gelfand--Tzetlin equations}
\label{sec24}

Up to the $k_1$--th level both sets of equations (\ref{gt50}) to
(\ref{gt52}) and (\ref{rec11}) to (\ref{rec13}) have to be solved for
even and odd levels separately. For the even levels, there is, as
already mentioned above, a direct correspondence to the case of the
unitary supergroup. Thus, we find employing the results of
Ref.~\onlinecite{GGT},
\begin{eqnarray}
|v_p^{(n)}|^2 &=&
            \frac{\prod_{q=1}^{\frac{k_1-n-1}{2}}
            \left((s_{p1}^{(n-1)})^2-(s_{q1}^{(n)})^2\right)
            \prod_{q=1}^{k_2}
            \left((s_{p1}^{(n-1)})^2-(is_{q2}^{(n-1)})^2\right)}
           {\prod_{q=1,q\neq p}^{\frac{k_1-n+1}{2}}
            \left((s_{p1}^{(n-1)})^2-(s_{q1}^{(n-1)})^2\right)
            \prod_{q=1}^{k_2}
            \left((s_{p1}^{(n-1)})^2-(is_{q2}^{(n)})^2\right)} \ ,
                                       \nonumber\\
  & & \qquad\qquad\qquad\qquad\qquad\qquad\qquad\qquad\qquad
    p=1,\ldots,(k_1-n+1)/2 \ , \quad k_1-n+1 \ {\rm even},
                                       \nonumber\\
|\alpha_p^{(n)}|^2 &=&
          \left((is_{p2}^{(n)})^2-(is_{p2}^{(n-1)})^2\right)
         \frac{\prod_{q=1}^{\frac{k_1-n-1}{2}}
                \left((is_{p2}^{(n-1)})^2-(s_{q1}^{(n)})^2\right)
                \prod_{q=1,q\neq p}^{k_2}
                \left((is_{p2}^{(n-1)})^2-(is_{q2}^{(n-1)})^2)\right)}
               {\prod_{q=1}^{\frac{k_1-n+1}{2}}
                \left((is_{p2}^{(n-1)})^2-(s_{q1}^{(n-1)})^2\right)
                \prod_{q=1,q\neq p}^{k_2}
  \left((is_{p2}^{(n-1)})^2-(is_{q2}^{(n)})^2\right)} \ , \nonumber\\
  & & \qquad\qquad\qquad\qquad\qquad\qquad\qquad\qquad\qquad
    p=1,\ldots,k_2 \ .
\label{Sol1}
\end{eqnarray}
We have included the first level by setting $s=s^{(0)}$.  To find the
solution of Eqs.~(\ref{rec11}) to~(\ref{rec13}), one cannot directly
make use of the results in the unitary case. An explicit calculation
is necessary which is given in App.~\ref{appA}. It yields
\begin{eqnarray}
|w_{p}^{(n)}|^2 &=& \frac{-\prod_{m=1}^{\frac{k_1-n+1}{2}}
                   \left((s_{m1}^{(n-1)})^2-(s_{p1}^{(n)})^2\right)
                   \prod_{q=1}^{k_2}
                   \left((s_{p1}^{(n)})^2-(is_{q2}^{(n)})^2\right)}
                   {2(s_{p1}^{(n)})^2
                   \prod_{q=1,q\neq p}^{\frac{k_1-n-1}{2}}
                   \left((s_{p1}^{(n)})^2-(s_{q1}^{(n)})^2\right)
                   \prod_{q=1}^{k_2}
   \left((s_{p1}^{(n)})^2-(is_{q2}^{(n-1)})^2\right)} \ , \quad
              p=1,\ldots,(k_1-n-1)/2 \ , \nonumber\\
|w_{\frac{k_1-n+1}{2}}^{(n)}|^2&=&
                  \frac{\prod_{m=1}^{\frac{k_1-n+1}{2}}(s_{m1}^{(n-1)})^2
                        \prod_{q=1}^{k_2}(is_{q2}^{(n)})^2}
             {\prod_{m=1}^{\frac{k_1-n-1}{2}}(s_{m1}^{(n)})^2
              \prod_{q=1}^{k_2}(is_{q2}^{(n-1)})^2} \ , \nonumber\\
|\beta_{p}^{(n)}|^2&=&
       \left((is_{p2}^{(n)})^2-(is_{p2}^{(n-1)})^2\right)
       \frac{\prod_{q=1,q\neq p}^{k_2}
                \left((is_{p2}^{(n)})^2-(is_{q2}^{(n)})^2\right)
                \prod_{q=1}^{\frac{k_1-n+1}{2}}
                \left((is_{p2}^{(n)})^2-(s_{q1}^{(n)})^2\right)}
               {2 (is_{p2}^{(n)})^2\prod_{q=1}^{\frac{k_1-n-1}{2}}
                \left((is_{p2}^{(n)})^2-(s_{q1}^{(n)})^2\right)
                \prod_{q=1,q\neq p}^{k_2}
                \left((is_{p2}^{(n)})^2-(is_{q2}^{(n-1)})^2\right)}
                                      \ , \nonumber\\
  & & \qquad\qquad\qquad\qquad\qquad\qquad\qquad\qquad\qquad
          p=1\ldots k_2
\label{Sol2}
\end{eqnarray}
We observe that the squares of the fermionic eigenvalues of the
different levels $(is_{p2}^{(n)})^2$ differ only by a nilpotent
variable. Hence, we introduce complex anticommuting variables
$\xi_p^{(n)}$ such that
\begin{equation}
|\xi_p^{(n)}|^2 = (is_{p2}^{(n)})^2-(is_{p2}^{(n-1)})^2 \ .
\label{Sol3}
\end{equation}
We emphasize that this feature is highly non--trivial: the difference
of the squared fermionic eigenvalues for two neighbouring levels can
be expressed as the modulus squared of one anticommuting variable.

The solutions of Eqs.~(\ref{rec7a}) to~(\ref{rec7}) for the odd
levels, i.e.~for $k_1-n+1$ odd, cannot directly be obtained by
adjusting the results of Ref.~\onlinecite{GGT}. However, as the
necessary modifications are intuitively clear, we do not derive the
solutions for the odd levels in detail. We simply state the results,
\begin{eqnarray}
|v_p^{(n)}|^2&=&\frac{\prod_{q=1}^{\frac{k_1-n}{2}}
               \left((s_{p1}^{(n-1)})^2-(s_{q1}^{(n)})^2\right)
               \prod_{q=1}^{k_2}
               \left((s_{p1}^{(n-1)})^2-(is_{q2}^{(n-1)})^2\right)}
               {(s_{p1}^{(n-1)})^2
               \prod_{q=1,q\neq p}^{\frac{k_1-n}{2}}
               \left((s_{p1}^{(n-1)})^2-(s_{q1}^{(n-1)})^2\right)
               \prod_{q=1}^{k_2}
      \left((s_{p1}^{(n-1)})^2-(is_{q2}^{(n)})^2\right)} \ ,
                             \quad p=1,\ldots,(k_1-n)/2 \ , \nonumber\\
|v_{\frac{k_1-n}{2}+1}^{(n)}|^2&=&
               \frac{\prod_{q=1}^{\frac{k_1-n}{2}}(s_{q1}^{(n)})^2
               \prod_{q=1}^{k_2}(is_{q2}^{(n-1)})^2}
              {\prod_{q=1}^{\frac{k_1-n}{2}}(s_{q1}^{(n-1)})^2
               \prod_{q=1}^{k_2}(is_{q2}^{(n)})^2} \ , \nonumber\\
|\alpha_p^{(n)}|^2&=&
             \left((is_{p2}^{(n)})^2-(is_{p2}^{(n-1)})^2\right)
             \frac{\prod_{q=1}^{\frac{k_1-n}{2}}
                    \left((is_{p2}^{(n-1)})^2-(s_{q1}^{(n)})^2\right)
                    \prod_{q=1,q\neq p}^{k_2}
             \left((is_{p2}^{(n-1)})^2-(is_{q2}^{(n-1)})^2)\right)}
                   {(is_{p2}^{(n-1)})^2
                    \prod_{q=1}^{\frac{k_1-n}{2}}
                    \left((is_{p2}^{(n-1)})^2-(s_{q1}^{(n-1)})^2\right)
                    \prod_{q=1,q\neq p}^{k_2}
    \left((is_{p2}^{(n-1)})^2-(is_{q2}^{(n)})^2\right)} \ ,\nonumber\\
  & & \qquad\qquad\qquad\qquad\qquad\qquad\qquad\qquad\qquad
            p=1,\ldots,k_2 \ .
\label{AB1}
\end{eqnarray}
The solutions of Eqs.~(\ref{rec11}) to~(\ref{rec13}) for the
odd levels read
\begin{eqnarray}
|w_{p}^{(n)}|^2&=&
             \frac{1}{2}
             \frac{\prod_{m=1}^{\frac{k_1-n}{2}}
                   \left((s_{m1}^{(n-1)})^2-(s_{p1}^{(n)})^2\right)
                   \prod_{q=1}^{k_2}
                   \left((s_{p1}^{(n)})^2-(is_{q2}^{(n)})^2\right)}
                   {\prod_{q=1,q\neq p}^{\frac{k_1-n}{2}}
                   \left((s_{p1}^{(n)})^2-(s_{q1}^{(n)})^2\right)
                   \prod_{q=1}^{k_2}
       \left((s_{p1}^{(n)})^2-(is_{q2}^{(n-1)})^2\right)} \ ,
                    \quad p=1,\ldots,(k_1-n)/2 \ , \nonumber\\
|\beta_{p}^{(n)}|^2&=& \frac{1}{2}
      \left((is_{p2}^{(n)})^2-(is_{p2}^{(n-1)})^2\right)
            \frac{\prod_{q=1,q\neq p}^{k_2}
                    \left((is_{p2}^{(n)})^2-(is_{q2}^{(n)})^2\right)
                    \prod_{q=1}^{\frac{k_1-n}{2}}
                    \left((is_{p2}^{(n)})^2-(s_{q1}^{(n-1)})^2\right)}
                   {\prod_{q=1}^{\frac{k_1-n}{2}}
                    \left((is_{p2}^{(n)})^2-(s_{q1}^{(n)})^2\right)
                    \prod_{q=1,q\neq p}^{k_2}
     \left((is_{p2}^{(n)})^2-(is_{q2}^{(n-1)})^2\right)} \ , \nonumber\\
  & & \qquad\qquad\qquad\qquad\qquad\qquad\qquad\qquad\qquad
         p=1,\ldots,k_2 \ .
\label{AB2}
\end{eqnarray}
{}From the solutions stated in Eqs.~(\ref{Sol1}) to (\ref{AB2}) one
derives the corresponding formulae for the group ${\rm SO}(k_1)$ in
ordinary space by setting all anticommuting variables to zero.

A comparison with the results for the unitary supergroup ${\rm
U}(k_1/2k_2)$ in Ref.~\onlinecite{GGT} reveals an interesting formal
connection. The Cartan subalgebras ${\rm u}^{(0)}(k_1-n+1/2k_2)$ and
${\rm u}^{(0)}(k_1-n/2k_2)$ of ${\rm U}(k_1-n+1/2k_2)$ and ${\rm
U}(k_1-n/2k_2)$, respectively, are all diagonal matrices
\begin{eqnarray}
s^{(n-1)}&=&\diag(s_{11}^{(n-1)},\ldots,s_{(k_1-n+1)1}^{(n-1)},
is_{12}^{(n-1)},\ldots,is_{2k_22}^{(n-1)}) \ , \nonumber\\
s^{(n)}&=&\diag(s_{11}^{(n)},\ldots,s_{(k_1-n)1}^{(n)},
is_{12}^{(n)},\ldots,is_{2k_22}^{(n)}) \ .
\label{Sol4}
\end{eqnarray}
If one now formally replaces, in the results for the unitary
supergroup, these matrices $s^{(n-1)}$ and $s^{(n)}$ with elements of
the Car\-tan sub\-alge\-bra of ${\rm uosp}^{(0)}(k_1-n+1/2k_2)$ and
${\rm uosp}^{(0)}(k_1-n/2k_2)$ according to
\begin{eqnarray}
  s^{(n-1)}&\longleftrightarrow&\diag(s_{11}^{(n-1)},\ldots,
  s_{\frac{k_1-n+1}{2}1}^{(n-1)},
  is_{12}^{(n-1)}\tau^{(3)},\ldots,is_{k_22}^{(n-1)})\otimes 
                i\tau^{(3)} \ ,
                               \nonumber\\
  s^{(n)}&\longleftrightarrow&\diag(s_{11}^{(n)},\ldots,
  s_{\frac{k_1-n-1}{2}1}^{(n-1)},0,
  is_{12}^{(n)},\ldots,is_{k_22}^{(n)})\otimes i\tau^{(3)} \ ,
                      \quad k_1-n+1 \ {\rm even}.
\label{Sol5}
\end{eqnarray}
the results in Eqs.~(\ref{Sol1}) and~(\ref{Sol2}) and in
Eqs.~(\ref{AB1}) and~(\ref{AB2}) are recovered. This formal connection
between the unitary supergroup and the unitary orthosymplectic one is
natural and plausible.  Unfortunately, we could not make a sound
mathematical reasoning out of the replacement~(\ref{Sol5}) which would
go beyond the {\it a posteriori} observation. However, the formal
connection stated above illustrates the deep relationship between the
groups which will become even more apparent in the generalized Gelfand
pattern given in Sec.~\ref{sec4}.

\subsection{Ordinary unitary symplectic group}
\label{sec25}

The $k_1$--th step of the recursion is the last one in a
superspace. We now approach the second line of Eq.~(\ref{gt0.1}).  The
following steps do not involve anticommuting variables anymore. We are
left with the ordinary unitary symplectic group ${\rm USp}(2k_2)$ and
its coset decomposition. We make use of the isomorphism ${\rm
USp}(2k_2)\cong {\rm U}(k_2;4)$ where ${\rm U}(k_2;4)$ is the unitary
group in $k_2$ dimensions parametrized over the quaternions. Since
${\rm U}(k_2;4)$ can be parametrized analogously to ${\rm
U}(k_2;2)\cong {\rm U}(k_2)$, i.e.~to the unitary group over the
complex numbers, we only have to adjust the results of
Refs.~\onlinecite{GT,SLS} where Gelfand--Tzetlin coordinates for the
ordinary unitary group were constructed. We write $U\in {\rm
U}(k_2;4)$ as $U=[U_1 \, U_2 \cdots U_{k_2}]$. The normalized vectors
$U_i$ have quaternionic entries.  The Cartan subalgebra is of the form
$s_2^{(k_1)}=\diag(s_{12}^{(k_1)},\ldots,s_{k_22}^{(k_1)})\otimes
\tau^{(3)}$.  The Gelfand--Tzetlin eigenvalue equation reads for the
first level of the ${\rm USp}(2k_2)$ recursion, i.e.~for the level
$k_1+1$ of the ${\rm UOSp}(k_1/2k_2)$ recursion
\begin{equation}
(1_{k_2}-U_1U_1^\dagger) is_2^{(k_1)} (1_{k_2}-U_1U_1^\dagger)
  E_n^{(1)} = is_{n2}^{(k_1+1)} E_n^{(1)} \ .
\label{sy1}
\end{equation}
We introduced capital letters $U_1=u_{k_1+1}$ and
$E_n^{(1)}=e_n^{(k_1+1)}$ in order to highlight that the vectors and
matrices used here live in ordinary space. Since the operator on the
left hand side is not Hermitean selfdual, Eq.~(\ref{sy1}) has not a
unique solution, see for example Ref.~\onlinecite{MEHTA2}. This can be
cured by multiplying Eq.~(\ref{sy1}) on both sides with
$1_{k_2}\otimes \tau^{(3)}$ from the right. A well defined eigenvalue
equation for a selfdual matrix obtains. It is known to have $k_2$
scalar eigenvalues $is_{p2}^{(k_1+1)}1_2$ which, to keep the notation
simple, we also denote by $is_{p2}^{(k_1+1)}$. After this adjustment,
we can proceed along the same lines which led to Eq.~(\ref{gt3}). The
equation reduces to the well known Gelfand--Tzetlin
equations~\cite{GT,SLS} of the unitary group ${\rm U}(k_2;2)\cong {\rm
U}(k_2)$,
\begin{eqnarray}
1 &=& \sum_{n=1}^{k_2} |U_{n1}|^2 \ ,\nonumber\\
0 &=& \sum_{m=1}^{k_2}
               \frac{|U_{m1}|^2}
              {is_{m2}^{(k_1)}-is_{p2}^{(k_1+1)}} \ ,
                               \qquad p=1,\ldots,k_2-1 \ .
\label{sy2}
\end{eqnarray}
This establishes a one--to--one correspondence between the $(k_2-1)$
eigenvalues $is_{p2}^{(k_1+1)}$ and the moduli squared of the
quaternionic entries
\begin{equation}
|U_{m1}|^2 = \Tr U_{m1}^\dagger U_{m1} \ . \label{sy3}
\end{equation}
All formulae derived in Refs.~\onlinecite{GT,SLS} for the unitary
group can now be adopted to the unitary symplectic one.

\subsection{Invariant measure}
\label{sec26}

According to the coset decomposition~(\ref{gt0.1}), the invariant
measure $d\mu(u)$ of $u \in {\rm UOSp}(k_1/2k_2)$ is the product of
all measures on the cosets, i.e.~on the spheres described by them. Of
course, these measures are conditioned, because the orthogonality of
the vectors $u_n$ in $u=[u_1 \, u_2 \cdots u_{k_1+2 k_2}]$ has to be
respected. As we will see, the Gelfand--Tzetlin coordinates take care
of this condition in a most convenient way. We evaluate the squared
invariant length element $du_n^\dagger du_n$. For $(k_1-n+1)$ even, it
reads,
\begin{eqnarray}
d{u_n^\prime}^\dagger du_n^\prime \ = \ du_n^\dagger
                   du_n&=&\sum_{m=1}^{\frac{k_1-n+1}{2}}
                    \frac{1}{4|v_m^{(n)}|^2}
                    \left(d|v_m^{(n)}|^2\right)^2+
                    \sum_{m=1}^{\frac{k_1-n+1}{2}}|v_m^{(n)}|^2
                    (d\vartheta_m^{(n)})^2+
                    \sum_{m^\prime=1}^{k_2}
                   d(\alpha_{m^\prime}^{(n)})^\ast
                   d\alpha_{m^\prime}^{(n)} \ .
\label{inv1}
\end{eqnarray}
where we use the parametrization~(\ref{gt1.1}). The first equality is
due to the basis independence of the invariant length element. It is a
highly welcome feature of the Gelfand--Tzetlin coordinates for the
unitary group in ordinary space~\cite{GT,SLS} and in
superspace~\cite{GGT} that the metric remains diagonal.  This holds
also in the present problem. Extending the corresponding calculation
of Ref.~\onlinecite{GGT}, we find for the even levels, i.e.~for
$k_1-n+1$ even,
\begin{eqnarray}
d\mu(u_n)=2^{k_2}\prod_{p=1}^{\frac{k_1-n-1}{2}}s_{p1}^{(n)}
                 \frac{B_{\frac{k_1-n-1}{2}k_2}\left((s^{(n)})^2\right)}
                {B_{\frac{k_1-n+1}{2}k_2}\left((s^{(n-1)})^2\right)}
              d[s_1^{(n)}]d[\vartheta^{(n)}]d[\xi^{(n)}] \ ,
         n\leq k_1 \ , \quad k_1-n+1 \ {\rm even} \ ,
\label{inv4}
\end{eqnarray}
and for the odd levels, i.e.~for $k_1-n+1$ odd, we have
\begin{eqnarray}
d\mu(u_n)=2^{k_2}\frac{\prod_{p=1}^{k_2}
                    is_{p2}^{(n)}is_{p2}^{(n-1)}}
                    {\prod_{p=1}^{\frac{k_1-n}{2}} s_{p1}^{(n-1)}}
                 \frac{B_{\frac{k_1-n}{2}k_2}\left((s^{(n)})^2\right)}
                {B_{\frac{k_1-n}{2}k_2}\left((s^{(n-1)})^2\right)}
                 d[s_1^{(n)}]d[\vartheta^{(n)}]d[\xi^{(n)}] \ ,
         n\leq k_1 \ , \quad k_1-n+1 \ {\rm odd}.
\label{inv7}
\end{eqnarray}
Here, we introduced the function
\begin{equation}
B_{nm}(s) = \frac{\prod_{p>q}^n(s_{p1}-s_{q1})
                  \prod_{p>q}^m(is_{p2}-is_{q2})}
                 {\prod_{p,q}(s_{p1}-is_{q2})}
          = \frac{\Delta_n(s_1)\Delta_m(is_2)}
                 {\prod_p^n\prod_q^m(s_{p1}-is_{q2})} \ .
\label{inv5}
\end{equation}
It contains the ordinary Vandermonde determinants $\Delta_n(s_1)$ and
$\Delta_m(is_2)$ and can be viewed as the supersymmetric
generalization of the Vandermonde
determinant~\cite{GGT,TG,GUH4}. Furthermore, we defined
\begin{equation}
   d[s_1^{(n)}] = \prod_{p=1}^{\frac{k_1-n+1}{2}} ds_{p1}^{(n)} \ , \quad
   d[\vartheta^{(n)}] = 
             \prod_{p=1}^{\frac{k_1-n+1}{2}} d\vartheta_p^{(n)}
   \quad {\rm and} \quad
   d[\xi^{(n)}] = \prod_{p=1}^{k_2} d\xi_p^{(n)*} d\xi_p^{(n)} \ .
\label{inv6}
\end{equation}
Remarkably, Eqs.~(\ref{inv4}) and~(\ref{inv7}) imply that the measures
on all cosets factorize.  Collecting all these measures up to the
$k_1$--th step, we obtain the invariant measure of $u\in{\rm
UOSp}(k_1/2k_2)$ in the form
\begin{equation}
d\mu(u)=2^{k_1k_2}\frac{\Delta_{k_2}\left((is_2^{(k_1)})^2\right)}
                  {B_{\frac{k_1}{2}k_2}\left(s^2\right)}
        \prod_{i=1}^{k_1}\prod_{p=1}^{k_2}is_{p2}^{(i)}
        d[s_1^{(i)}]d[\vartheta^{(i)}]d[\xi^{(i)}]
        d\mu(U) \ ,
\label{inv8}
\end{equation}
where $d\mu(U)$ is the invariant measure on $U\in{\rm USp}(2k_2)$. We
mention in passing that the measure of the orthogonal group in
ordinary space can be obtained by setting all anticommuting variables
to zero in the invariant length~(\ref{inv1}), and skipping all
couplings between the bosonic and fermionic eigenvalues in
Eq.~(\ref{inv8}).

The measure~(\ref{inv8}) on ${\rm UOSp}(k_1/2k_2)$ has an important
feature: Most conveniently, it is, apart from $d\mu(U)$, flat. This
follows directly from the factorization of the measures~(\ref{inv4})
and~(\ref{inv7}) on the coset spaces.  This is also true for the
Gelfand--Tzetlin coordinates of the unitary group in
ordinary~\cite{GT,SLS} and in superspace~\cite{GGT} as well as for the
ones of the orthogonal group in ordinary space.  However, this
important feature does not continue beyond the $k_1$--th level.  We
will see that now in working out the measure $d\mu(U)$ for $U\in{\rm
USp}(2k_2)$.  We write $u=[U_1 \, U_2 \cdots U_{k_2}]$ and decompose
the entry $U_{nm}$ as $U_{nm}=|U_{nm}|\widehat{U}_{nm}$, with
$\widehat{U}_{nm}$ a unimodular quaternion. We introduce a
parametrization of the unimodular quaternion,
\begin{equation}
\widehat{U}_{nm}=\left[\matrix{
                 \cos \psi_n^{(m)} \exp(-i\gamma_{n1}^{(m)})&
                -\sin \psi_n^{(m)} \exp(i\gamma_{n2}^{(m)})\cr
                 \sin \psi_n^{(m)} \exp(-i\gamma_{n2}^{(m)})&
                 \cos \psi_n^{(m)} \exp(i\gamma_{n1}^{(m)})}\right]
                 \ ,
\label{sy5}
\end{equation}
which allows us to write the invariant length element squared as
\begin{equation}
\Tr dU_m^\dagger dU_m = \sum_{n=1}^{k_2}
                          \left(\frac{1}{4|U_{nm}|^2}(d|U_{nm}|^2)^2
                          +\sum_{i=1}^{2}|U_{nm}|^2(d\gamma_{ni}^{(m)})^2
                          +|U_{nm}|^2(d\cos\psi_n^{(m)})^2\right)\ ,
                          \ m=1\ldots k_2\ .
\label{sy6}
\end{equation}
Employing and properly adjusting the results of
Refs.~\onlinecite{GT,SLS}, one finds the measure of the coset in the
$m$--th level
\begin{equation}
d\mu(U_m) = \frac{\Delta_{k_2-m}(is_2^{(k_1+m)})}
                 {2^{k_2-m}\Delta_{k_2-m+1}^3(is_2^{(k_1+m-1)})}
            \prod_{p,q}(is_{p2}^{(k_1+m-1)}-is_{q2}^{(k_1+m)})
        d[s_2^{(k_1+m)}]d[\cos\psi^{(m)}]d[\gamma^{(m)}] \ ,
\label{sy9}
\end{equation}
with the definitions
\begin{equation}
   d[s_2^{(k_1+m)}] = \prod_{p=1}^{k_2-m+1} ds_{p2}^{(k_1+m)} \ , \quad
   d[\cos\psi^{(m)}] = \prod_{p=1}^{k_2-m+1} d\cos\psi_p^{(m)}
   \quad {\rm and} \quad
   d[\gamma^{(m)}] = \prod_{p=1}^{k_2-m+1}
                 d\gamma_{p1}^{(m)}d\gamma_{p2}^{(m)}\ .
\label{sy9a}
\end{equation}
One clearly sees that the factorization property does not hold for the
unitary symplectic group. This is a peculiarity of the
Gelfand--Tzetlin parametrization for the unitary symplectic
group. Collecting all levels we arrive at the invariant measure on
$U\in{\rm USp}(2k_2)$,
\begin{equation}
d\mu(U) =\frac{1}{2^{k_2(k_2-1)/2}\Delta_{k_2}^3(is_2^{(k_1)})}
         \prod_{m=1}^{k_2}
         \prod_{n=1}^{k_2-m+1} \prod_{n^\prime=1}^{k_2-m}
         \frac{is_{n2}^{(k_1+m-1)}-is_{n^\prime2}^{(k_1+m)}}
         {\Delta_{k_2-m}^2(is_2^{(k_1+m)})}
         d[s_2^{k_1+m}]d[\cos\psi^{(m)}]d[\gamma^{(m)}]
\label{sy10}
\end{equation}
which combines with Eq.~(\ref{inv8}) to the full measure on ${\rm
UOSp}(k_1/2k_2)$.

\subsection{Matrix elements}
\label{sec27}

With the results of the previous sections, we can express an arbitrary
column $u_p$ of a matrix $u=[u_1 \, u_2 \cdots u_{k_1+2 k_2}]$ in the
unitary orthosymplectic supergroup ${\rm UOSp}(k_1/2k_2)$ in terms of
our angular Gelfand--Tzetlin coordinates. In the rotated primed
basis~(\ref{gt2}), we have
\begin{equation}
u_p^\prime=\mbox{$\widehat{b}^{\prime(1)}$}^T
           \mbox{$\widehat{b}^{\prime(2)}$}^T\ldots
           \mbox{$\widehat{b}^{\prime(n-1)}$}^T u_p^{\prime(n-1)}\ ,
\label{me1}
\end{equation}
where $b^{(n)}$ and the scalar products are defined in
Eq.~(\ref{rec2}) and~(\ref{rec3}) and in Eqs.~(\ref{rec8})
and~(\ref{rec10}). So far, we have constructed a unitary
representation of ${\rm UOSp}(k_1/2k_2)$. We also wish to obtain an
orthosymplectic representation. To this end, we have to assure that
the vectors $u_j,\ j\leq k_1$ become real, when the matrix
$u^{\prime}=\left[u_1^\prime \, u_2^\prime \cdots
u_{k_1+k_2}^\prime\right]$ is rotated back into the unprimed basis. We
only discuss the case $k_1-n+1$ even, the odd case is treated
analogously. We recall that the vector $b^{(n)}$ entering in the
projection matrix in Eq.~(\ref{rec10}) has been determined only up to
a phase.  There is an ambiguity in choosing the phase of $b^{(n)}$.
The Gelfand--Tzetlin coordinates parametrize the vector $u_n$ only up
to some phases associated with the action of the Cartan subgroup of
${\rm UOSp}\left((k_1-n+1)/2k_2\right)$.  Thus, the projection matrix
$\widehat{b}^{(n)}$ is as well invariant under the action of this
Cartan subgroup. We may multiply $\widehat{b}^{(n)}$ with an arbitrary
element of the Cartan subgroup without changing its projection
properties. We set $b^{(n)}_{2p}=ib^{(n)}_{2p-1}, \ p\leq (k_1-n-1)/2$
and $b^{(n)}_{k_1-n}=i |w_{\frac{k_1-n+1}{2}}|$ in the commuting
sector and $b^{(n)}_{k_1-n+1+2p}=-b^{(n)*}_{k_1-n+2p}, \
p=1,\ldots,k_2$ in the anticommuting one. The remaining phases may be
set to zero.  With this choice of phases and after undoing the basis
rotation, the columns as well as the rows of $\widehat{b}^{(n)T}$
fulfill the reality condition~(\ref{gt0}).  The vectors $u_n^{(n-1)}$
become real, too.  An explicit form of the real matrices
$\widehat{b}^{(n)}$ is given in App.~\ref{appB}.

%\section{Supersymmetric Harish-Chandra formula}
%\label{sec2.6}
%The Harish Chandra Integral is of the form
%\begin{equation}
%I(g,h) = \int \exp\left(i\trg u^{-1}hug\right) d\mu(u) \ ,
%\end{equation}{HC1}
%where $u\in{\rm UOSp}(k_1/2k_2)$ and $h,g$ are two elements of the
%Cartan Subalgebra. We firstly discuss the case, that $k_1$ is even.
%If we consider the quadratic form
%$\eta= g^{-1}hu$, we need obviously only $\eta_{n(n+1)}$,
%for $n\leq k_1$ odd and $\eta_{nn}$ for $k_1<n\leq 2k_2$.
%A short calculation yields
%\begin{eqnarray}
%\eta_{n(n+1)}& = & i\left(\sum_q^{\frac{k_1-n-1}{2}}
%                    \sqrt{2}|w_q^{(n)}||v_q^{(n+1)}|
%                    \sin\vartheta_q^{(n+1)}\right.\cr
%               &   &\left.-i\sum_q^{k_2}\frac{1}{\sqrt{2}}
%                    \left(\beta_q^{(n)*}\alpha_q^{(n+1)}
%                    +\beta_q^{(n)}\alpha_q^{(n+1)*}\right)
%                    +|w_{\frac{k_1-n+1}{2}}^{(n)}|
%                    |v_{\frac{k_1-n+1}{2}}^{(n+1)}|\right)\cr
%               &   &\qquad\qquad\qquad n=1,\ldots,k_1-1\cr
%\eta_{(k_1+2n)(k_1+2n)} & = &-\eta_{(k_1+2n-1)(k_1+2n-1)}\cr
%            & = &\sum_{i=1}^{k_2-n+1}
%                i|v_i^{(k_1+n)}|^2 s_{i2}^{(k_1+n-1)}\cos 2\gamma_i\cr
%            &   &  n=1,\ldots,k_2
%\end{eqnarray}{HC2}

\section{Generalized Gelfand pattern}
\label{sec3}

The unitary Lie group ${\rm U}(k;\beta)$ over the real ($\beta=1$) and
complex ($\beta=2$) numbers and over the quaternions ($\beta=4$) is
isomorphic to the orthogonal, unitary and unitary symplectic group,
${\rm SO}(k)\cong {\rm U}(k;1)$, ${\rm U}(k)\cong {\rm U}(k;2)$ and
${\rm USp}(2k)\cong {\rm U}(k;4)$.  The Gelfand--Tzetlin
representation scheme obtains from the following procedure~\cite{BR}.
An irreducible representation is defined by an ordered set of integers
or half integers called highest weights. This irreducible
representation can be decomposed in irreducible representations of
${\rm U}(k-1;\beta)$. In the decomposition each irreducible
representation of ${\rm U}(k-1;\beta)$ occurs either exactly once or
never. Only those irreducible representations appear whose highest
weights satisfy certain betweenness conditions depending on the group
under consideration. Going through all steps of the group chain or,
equivalently, the coset decomposition,
\begin{equation}
{\rm U}(k;\beta)=
      \frac{{\rm U}(k;\beta)}{{\rm U}(k-1;\beta)}\otimes
      \frac{{\rm U}(k-1;\beta)}{{\rm U}(k-2;\beta)}\otimes
      \cdots \otimes
      \frac{{\rm U}(2;\beta)}{{\rm U}(1;\beta)}\otimes
      {\rm U}(1;\beta) \ ,
\label{gt01}
\end{equation}
one has labelled all states in the irreducible representation of ${\rm
U}(k;\beta)$ by a set of integers or half integers, arranged in a
Gelfand pattern.

The analogue for the coordinates is as follows. We consider the
adjoint group action $\mathcal{O}_k =U^\dagger x U$ on an element $x$
of the Cartan subalgebra ${\rm u}^{(0)}(k;\beta)$ with $U\in {\rm
U}(k;\beta)$. Here, in this one instance, we use the symbol $x$ for an
element in the algebra, because we want to emphasize that the present
discussion so far applies to ordinary groups and because we want to
avoid confusion with the discussion to follow on the supergroups. This
subset $\mathcal{O}_k =U^\dagger x U$ of the complete algebra is
called {\it orbit}. We can map the ${\rm U}(k;\beta)$ orbit labelled
by an ordered set of eigenvalues $x_i > x_{i+1}$ onto many different
${\rm U}(k-1;\beta)$ orbits by projecting $\mathcal{O}_k$ onto a $k-1$
dimensional subspace. But only those ${\rm U}(k-1;\beta)$ orbits
$\mathcal{O}_{k-1}$ can be reached, whose eigenvalues interlace two
neighboring eigenvalues of $\mathcal{O}_k$. This is the so called
minimax principle for selfadjoint operators~\cite{DS}. The
Gelfand--Tzetlin method uses the eigenvalues of the projected matrix
as coordinates of the coset ${\rm U}(k;\beta)/{\rm
U}(k-1;\beta)$. However, $x$ is a fixed point of the action of the
Cartan subgroup $\exp\left(i x_0\right), \ x_0 \in {\rm
u}^{(0)}(k;\beta)$.  Hence, the coset ${\rm U}(k;\beta)/{\rm
U}(k-1;\beta)$ is parametrized by the eigenvalues of
$\mathcal{O}_{k-1}$ only up to equivalence classes with respect to the
action of the Cartan subgroup of ${\rm U}(k;\beta)$, parametrized by
$x_0$. In this way the set of variables describing the coset is split
into two parts: One part consists of the eigenvalues of
$\mathcal{O}_{k-1}$, the other one of the independent elements of
$x_0$. Guillemin and Sternberg~\cite{GS1} introduced the concept of
complete integrability by interpreting the entries of $x$ as action
and the elements of $x_0$ as angle coordinates of a generalized
mechanical system. We emphasize that this usage of the term angles is
different from the one introduced previously. We distinguish between
radial and angular Gelfand--Tzetlin coordinates. However, both
Gelfand--Tzetlin coordinates, the radial and the angular ones, allow
for a further distinction between action and angle degrees of freedom,
although the interpretation is slightly different in the two cases.
The Guillemin--Sternberg theory applies only to the groups ${\rm
U}(k;\beta)$ for $\beta=1,2$ but not to the unitary symplectic
group. This can be considered as the reason for the relatively
complicated expression of the measure for ${\rm USp}(2k)\cong {\rm
U}(k;4)$.

The generalized Gelfand pattern for${\rm UOSp}(k_1/2k_2)$ can be
extracted from the positive definiteness of the moduli squared of the
bosonic matrix elements $|v_i^{(n)}|^2$. If one restricts oneself to
the subgroup, which consists of the direct product ${\rm
SO}(k_1)\otimes {\rm USp}(2k_2)$ the pattern of the ${\rm SO}(k_1)$
and ${\rm USp}(2k_2)$ are rederived which are well known from
representation theory~\cite{BR}.  We state them here in a different
form which emphasizes the relation to the pattern of the unitary group
${\rm U}(k)$ which is the famous triangle~\cite{GN}
\begin{eqnarray}
& \matrix{
\ul{x_1^{(0)}}&&\ul{x_2^{(0)}}&&\cdots &&\ul{x_{k}^{(0)}}&&\ul{x_{k+1}^{(0)}} \cr
        & x_1^{(1)}&&x_2^{(1)}&\cdots & x_{k-2}^{(1)}&&x_{k}^{(1)}& \cr
            & &    &          & \vdots & &    &        & & \cr
               & & &x_1^{(k-1)} & & x_2^{(k-1)} & & & &\cr
                    & & & & x_1^{(k)} & & & &
} &
\label{pat1}
\end{eqnarray}
with the betweeness conditions
\begin{equation}
x_{i+1}^{(j-1)}\leq x_i^{(j)}\leq x_i^{(j-1)} \ .
\label{pat1a}
\end{equation}
The first row in the pattern~(\ref{pat1}) labels the orbit which was
used as the starting point for the construction of the
parametrization. We underline them to distinguish them from the
coordinates of the group.  From this pattern the pattern of the
orthogonal group can be derived by the substitution rule~(\ref{Sol5}),
i.e.~by assigning to the Cartan subalgebra ${\rm u}^{(0)}(k)$ of the
unitary group ${\rm U}(k)$ the corresponding one ${\rm so}^{(0)}(k)$
of the orthogonal group ${\rm SO}(k)$. We restrict ourselves to the
case of even $k$. The pattern (\ref{pat1}) acquires the form
\begin{eqnarray}
\begin{array}{ccccccccccccccc}
 \ul{+\xu{1}{0}}&
       &\ul{+\xu{2}{0}}&&\ldots&&\ul{+\xu{k+1}{0}}&
&\ul{-\xu{k+1}{0}}&&\ldots&&\ul{-\xu{2}{0}}&&\ul{-\xu{1}{0}}\cr
&\xu{1}{1}&&\xu{2}{1}&\ldots&\xu{k}{1}&&0&
&\ul{-\xu{k}{1}}&\ldots&\ul{-\xu{2}{1}}&&\ul{-\xu{1}{1}}&\cr
&&&&&&& \vdots &&&&&&& \cr
&&&&\xu{1}{2k-2}&&\xu{2}{2k-2}&&\ul{-\xu{2}{2k-2}}&&\ul{-\xu{1}{2k-2}}&&&&\cr
&&&&&\xu{1}{2k-1}&&0&&\ul{-\xu{1}{2k-1}}&&&&&\cr
&&&&&&\xu{1}{2k}&&\ul{-\xu{1}{2k}}&&&&&&\cr
&&&&&&& 0  &&&&&&&
\end{array}
\label{pat2}
\end{eqnarray}
with the betweeness conditions
\begin{eqnarray}
x_{i+1}^{(j-1)}&\leq x_i^{(j)}\leq &x_i^{(j-1)}\cr
|x_{j}^{(2k-2j+2)}|&\leq &x_{j}^{(2k-2j+1)} \ .
\label{pat3}
\end{eqnarray}
We notice the symmetry along the middle axis. The variable space of
the ${\rm SO}(2k+2)$ is already covered by the left half of the
triangle. The other half is shown to indicate its relation to the
unitary case~(\ref{pat1}). Physically, this symmetry is due to time
reversal invariance: A system which is not invariant under time
reversal is modelled by hermitean operators. One can go to a time
reversal invariant system by replacing these operators with real
symmetric ones.  Restricting the pattern~(\ref{pat2}) to the left half
of the triangle, the patterns appear in their traditional
form~\cite{GT2}. By construction, the pattern of the unitary
symplectic group ${\rm USp}(2k)$ coincides with the one of the unitary
group ${\rm U}(k)$, this is due to the fact that only the action
variables are used in the pattern. The unitary group has only one, but
the unitary symplectic group has three angle variables coming with
every action.

The two patterns of the orthogonal and the unitary symplectic groups
together represent the subgroup ${\rm SO}(k_1)\otimes {\rm USp}(2k_2)$
of ${\rm UOSp}(k_1/2k_2)$. What represents the coset ${\rm
UOSp}(k_1/2k_2)/({\rm SO}(k_1)\otimes {\rm USp}(2k_2))$ ? -- We observe
that the lengths squared $|\xi_p^{(n)}|^2$ of the anticommuting
variables $\xi_p^{(n)}$ introduced in Eq.~(\ref{Sol3}) have a
distinguished meaning.  We may identify these lengths of the
anticommuting variables as the analogues of the actions stemming from
the commuting degrees of freedom. We can organize the lengths squared
$|\xi_p^{(n)}|^2$ in a rectangular pattern.  Thus, the generalized
Gelfand pattern for the unitary orthosymplectic supergroup ${\rm
UOSp}(k_1/2k_2)$ obtains,
\begin{eqnarray}
&\begin{array}{ccccccccccccccc}
 \ul{+\su{1}{0}}&
       &\ul{+\su{2}{0}}&&\ldots&&\ul{+\su{k+1}{0}}&
&\ul{-\su{k+1}{0}}&&\ldots&&\ul{-\su{2}{0}}&&\ul{-\su{1}{0}}\cr
&\su{1}{1}&&\su{2}{1}&\ldots&\su{k}{1}&&0&
&\ul{-\su{k}{1}}&\ldots&\ul{-\su{2}{1}}&&\ul{-\su{1}{1}}&\cr
&&&&&&& \vdots &&&&&&& \cr
&&&&\su{1}{2k-2}&&\su{2}{2k-2}&&\ul{-\su{2}{2k-2}}&&\ul{-\su{1}{2k-2}}&&&&\cr
&&&&&\su{1}{2k-1}&&0&&\ul{-\su{1}{2k-1}}&&&&&\cr
&&&&&&\su{1}{2k}&&\ul{-\su{1}{2k}}&&&&&&\cr
&&&&&&& 0  &&&&&&&
\end{array}&\cr
\nonumber\\
&\matrix{
|\xi_1^{(1)}|^2 & |\xi_2^{(1)}|^2 & & &
     \cdots & & & |\xi_{k_2-1}^{(1)}|^2 & |\xi_{k_2}^{(1)}|^2 \cr
|\xi_1^{(2)}|^2 & |\xi_2^{(2)}|^2 & & &
     \cdots & & & |\xi_{k_2-1}^{(2)}|^2 & |\xi_{k_2}^{(2)}|^2 \cr
& & & & \vdots & & & & \cr
|\xi_1^{(k_1)}|^2 & |\xi_2^{(k_1)}|^2 & & &
     \cdots & & & |\xi_{k_2-1}^{(k_1)}|^2 & |\xi_{k_2}^{(k_1)}|^2
}\hspace{1.1cm}& \cr
\nonumber\\
&&\cr
&\matrix{ s_{12}^{(k_1)}&&s_{22}^{(k_1)}&&\cdots&
&s_{(k_2-1)2}^{(k_1)}&&s_{k_22}^{(k_1)} \cr
&s_{12}^{(k_1+1)}&&s_{22}^{(k_1+1)}&\cdots
&s_{(k_2-2)2}^{(k_1+1)}&&s_{(k_2-1)2}^{(k_1+1)}& \cr
&&&& \vdots &&&&& \cr & &
&s_{12}^{(k_1+k_2-1)} & & s_{22}^{(k_1+k_2-1)}&&&&\cr
& & & & s_1^{(k_1+k_2)} & & & &}\hspace{0.2cm} &
\label{pat4}
\end{eqnarray}
with the betweeness conditions
\begin{eqnarray}
s_{(i+1)1}^{(m-1)}&\leq s_{i1}^{(m)}\leq& s_{i1}^{(m-1)}\cr
s_{(i+1)2}^{(k_1+l)}&\leq s_{i2}^{(k_1+l+1)}\leq& s_{i2}^{(k_1+l)}\cr
-s_{j1}^{(k_1-2j-1)}&\leq s_{j1}^{(k_1-2j)} \leq& s_{j1}^{(k_1-2j-1)} ,
\label{bound}
\end{eqnarray}
where $1 \leq j \leq k_1/2-1$, $1 \leq m \leq k_1-2$ and $0\leq l \leq
k_2-1$. It was shown in Ref.~\onlinecite{GUH5} that the unitary
supergroup ${\rm U}(1/1)$ can be represented by supersymmetric
generalizations of Wigner functions. This representation of the
supergroup ${\rm U}(1/1)$ is labelled by the length of an
anticommuting variable. Therefore, we want to interpret the
generalized Gelfand pattern (\ref{pat4}) as labelling another kind of
representation which involves anticommuting variables as labels. The
two triangles label the basis of an irreducible representation of the
product ${\rm SO}(k_1)\otimes {\rm USp}(2k_2)$, whereas the remaining
coset ${\rm UOSp}(k_1/2k_2)/\left({\rm SO}(k_1)\otimes {\rm
USp}(2k_2)\right)$ is represented by the rectangular block of the
lengths squared of anticommuting variables. This extends the
corresponding considerations for the unitary supergroup in
Ref.~\onlinecite{GGT}. It is challenging to find a further
interpretation of these new representations of supergroups, possibly
by generalizing the Guillemin--Sternberg theory.

\section{Summary and conclusions}
\label{sec4}

We constructed Gelfand--Tzetlin coordinates for the unitary
orthosymplectic supergroup ${\rm UOSp}(k_1/2k_2)$. To this end, we
further extended the construction for the unitary supergroup ${\rm
U}(k_1/k_2)$. We obtained {\it angular} Gelfand--Tzetlin coordinates,
which always live in the space of the unitary orthosymplectic
supergroup. They ought to be distinguished from {\it radial}
Gelfand--Tzetlin which map group degrees of freedom onto those of
another space. We also calculated the invariant Haar measure on ${\rm
UOSp}(k_1/2k_2)$ and obtained an expression that is fairly simple due
to the recursive structure of the coordinates.  As the orthogonal and
the unitary symplectic groups are subgroups of the unitary
orthosymplectic supergroup, our construction also includes {\it
angular} Gelfand--Tzetlin coordinates on these ordinary groups.

The Gelfand--Tzetlin coordinates can be arranged in a generalized
Gelfand pattern.  A remarkable feature of this pattern is the
appearance of moduli squared of anticommuting variables. We argued
that an interpretation of these anticommuting variables as eigenvalues
of a set of invariant operators is likely to exist. It is an
interesting task to clarify the r\^ole of these anticommuting
variables in the representation theory for supergroups.

So far, Gelfand--Tzetlin coordinates were only constructed for compact
groups. But there is no apparent obstacle to construct them also for
non--compact groups. It would be interesting to see if such a
construction is indeed possible and how the non--compactness of some
variables is reflected in the corresponding Gelfand pattern, in
ordinary and in superspace.

\begin{acknowledgments}
TG and HK acknowledge financial support from the Swedish Research
Council and from the RNT Network of the European Union with Grant
No.~HPRN--CT--2000-00144, respectively. HK also thanks the division of
Mathematical Physics, LTH, for its hospitality during his visits to
Lund.
\end{acknowledgments}

\appendix

\section{Solution of Eqs.~(\ref{rec11}) to~(\ref{rec13})}
\label{appA}

We consider Eq.~(\ref{rec11}) and insert the solutions for
$|v_m^{(n)}|^2$ and $|\alpha_m^{(n)}|^2$ given in (\ref{Sol1}). The
right hand side of Eq.~(\ref{rec11}) can then be expanded in a sum of
monomials in the nilpotent Gelfand--Tzetlin variables
$|\xi_q^{(n)}|^2, \ q=1,\ldots,k_2$. Since each of the
$|\xi_q^{(n)}|^2$ only appears linearly, the rank of the monomials
cannot exceed $k_2$. Thus, we can rewrite Eq.~(\ref{rec11}) in
the form
\begin{equation}
\frac{1}{|w_{p}^{(n)}|^2}= \sum_{r=0}^{k_2} M^{(r)} \ ,
\label{A1}
\end{equation}
Where $M^{(r)}$ is the nilpotent part of $1/|w_{p}^{(n)}|^2$,
consisting of monomials in $|\xi_q^{(n)}|^2$ with rank $r$.
Explicitly we have
\begin{eqnarray}
M^{(r)}&=&\sum_{j_1\leq j_2\leq\ldots\leq j_r}^{k_2}
          \sum_m^{\frac{k_1-n+1}{2}}
          \frac{\prod_{q=1,q\neq p}^{\frac{k_1-n-1}{2}}
          \left((s_{m1}^{(n-1)})^2-(s_{q1}^{(n)})^2\right)}
         {\prod_{q=1,q\neq m}^{\frac{k_1-n+1}{2}}
          \left((s_{m1}^{(n-1)})^2-(s_{q1}^{(n-1)})^2\right)}
          \frac{(s_{m1}^{(n-1)})^2+(s_{p1}^{(n)})^2}
          {(s_{m1}^{(n-1)})^2-(s_{p1}^{(n)})^2}
          \frac{\prod_{i=1}^r |\xi_{j_i}|^2}
  {\prod_{i=1}^r\left((s_{m1}^{(n-1)})^2-(is_{j_i2}^{(n-1)})^2\right)}
                                           \nonumber\\
       & &+\sum_{\stackrel{j_2\leq\ldots\leq j_r}{j_1\neq j_i}}^{k_2}
          \frac{\prod_{q=1,q\neq p}^{\frac{k_1-n-1}{2}}
          \left((is_{j_12}^{(n-1)})^2-(s_{q1}^{(n)})^2\right)}
         {\prod_{q=1}^{\frac{k_1-n+1}{2}}
          \left((is_{j_12}^{(n-1)})^2-(s_{q1}^{(n-1)})^2\right)}
          \frac{(is_{j_12}^{(n-1)})^2+(s_{p1}^{(n)})^2}
          {(is_{j_12}^{(n-1)})^2-(s_{p1}^{(n)})^2}
       \frac{\prod_{i=1}^r |\xi_{j_i}|^2}
 {\prod_{i=2}^r\left((is_{j_12}^{(n-1)})^2-(is_{j_i2}^{(n-1)})^2\right)}
                                              \ .
\label{A2}
\end{eqnarray}
for $r=1,\ldots,k_2$.  The sum over $m$ is the Laplace expansion of a
determinant. For its evaluation we use the formula
\begin{eqnarray}
 \frac{1}{\prod_{i=1}^{r}
  \left((s_{m1}^{(n-1)})^2-(is_{i2}^{(n-1)})^2\right)}=
        \sum_{i=1}^r\frac{1}{(s_{m1}^{(n-1)})^2-(is_{i2}^{(n-1)})^2}
        \frac{1}{\prod_{i^\prime \neq i}^r
   \left((is_{i2}^{(n-1)})^2-(is_{i^\prime2}^{(n-1)})^2\right)} \ ,
\label{A3}
\end{eqnarray}
which is well known from complex analysis. After symmetrizing the
second sum in the indices, $j_i,i=1,\ldots,r$, we arrive at the
following expression for $M^{(r)}$.
\begin{eqnarray}
M^{(r)}&=&\sum_{j_1\leq j_2\leq\ldots\leq j_r}^{k_2}\sum_{i=1}^r
         \frac{1}
        {\prod_{i\prime \neq i}^r
  \left((is_{j_i2}^{(n-1)})^2-(is_{j_{i^\prime}2}^{(n-1)})^2\right)}
          \sum_m^{\frac{k_1-n+1}{2}}
         \frac{\prod_{q=1,q\neq p}^{\frac{k_1-n-1}{2}}
         \left((s_{m1}^{(n-1)})^2-(s_{q1}^{(n)})^2\right)}
        {\prod_{q=1,q\neq m}^{\frac{k_1-n+1}{2}}
         \left((s_{m1}^{(n-1)})^2-(s_{q1}^{(n-1)})^2\right)}\nonumber\\
     & & \left[\frac{(s_{m1}^{(n-1)})^2+(s_{p1}^{(n)})^2}
        {\left((s_{p1}^{(n)})^2-(is_{j_i2}^{(n-1)})^2\right)
         \left((s_{m1}^{(n-1)})^2-(s_{p1}^{(n)})^2\right)}-
         \frac{(s_{m1}^{(n-1)})^2+(s_{p1}^{(n)})^2}
        {\left((s_{p1}^{(n)})^2-(is_{j_i2}^{(n-1)})^2\right)
         \left((s_{m1}^{(n-1)})^2-(is_{j_i2}^{(n-1)})^2\right)}\right]
         \prod_{i=1}^r|\xi_{j_i}|^2\nonumber\\
     &+&\sum_{j_1\leq j_2\leq\ldots\leq j_r}^{k_2}\sum_{i=1}^r
         \frac{1}{\prod_{i\prime \neq i}^r
    \left((is_{j_i2}^{(n-1)})^2-(is_{j_{i^\prime}2}^{(n-1)})^2\right)}
                            \nonumber\\
     & & \qquad\qquad\qquad\qquad\qquad
         \frac{\prod_{q=1,q\neq p}^{\frac{k_1-n-1}{2}}
         \left((is_{j_12}^{(n-1)})^2-(s_{q1}^{(n)})^2\right)}
        {\prod_{q=1}^{\frac{k_1-n+1}{2}}
         \left((is_{j_12}^{(n-1)})^2-(s_{q1}^{(n-1)})^2\right)}
         \frac{(is_{j_12}^{(n-1)})^2+(s_{p1}^{(n)})^2}
        {(is_{j_12}^{(n-1)})^2-(s_{p1}^{(n)})^2}
         \prod_{i=1}^r|\xi_{j_i}|^2 \ .
\label{A4}
\end{eqnarray}
Now the determinant mentioned above can be evaluated by using the
translational invariance of the differences.  The second term in the
squared bracket cancels completely, we are left with
\begin{eqnarray}
M^{(r)}&=&\sum_{j_1\leq j_2\leq\ldots\leq j_r}^{k_2}\sum_{i=1}^r
          \frac{1}
          {\prod_{i\prime \neq i}^r
  \left((is_{j_i2}^{(n-1)})^2-(is_{j_{i^\prime}2}^{(n-1)})^2\right)}
                           \nonumber\\
  & & \qquad\qquad\qquad\qquad\qquad
  \frac{2(s_{p1}^{(n)})^2\prod_{q=1,q\neq p}^{\frac{k_1-n-1}{2}}
            \left((s_{q1}^{(n)})^2-(s_{p1}^{(n)})^2\right)}
           {\left((s_{p1}^{(n)})^2-(is_{j_i2}^{(n-1)})^2\right)
            \prod_{m=1}^{\frac{k_1-n+1}{2}}
            \left((s_{m1}^{(n-1)})^2-(s_{p1}^{(n)})^2\right)}
            \prod_{i=1}^r|\xi_{j_i}|^2 \ .
\label{A5}
\end{eqnarray}
Using identity (\ref{A3}) once more and summing over $r$ gives
\begin{eqnarray}
\frac{1}{|w_{p}^{(n)}|^2}=
                  \frac{-2(s_{p1}^{(n)})^2
                  \prod_{q=1,q\neq p}^{\frac{k_1-n-1}{2}}
                  \left((s_{p1}^{(n)})^2-(s_{q1}^{(n)})^2\right)}
                 {\prod_{m=1}^{\frac{k_1-n+1}{2}}
                  \left((s_{m1}^{(n-1)})^2-(s_{p1}^{(n)})^2\right)}
   \left(\sum_{r=0}^{k_2}\sum_{j_1\leq j_2\leq\ldots\leq j_r}^{k_2}
      \prod_{i=1}^r
 \frac{|\xi_{j_i}|^2}{(s_{p1}^{(n)})^2-(is_{j_i2}^{(n-1)})^2}\right) \ .
\label{A6}
\end{eqnarray}
The double sum in Eq.~(\ref{A6}) simply amounts to
\begin{equation}
\prod_{q=1}^{k_2}
\left(1+\frac{|\xi_q^{(n)}|^2}
      {(s_{p1}^{(n)})^2-(is_{q2}^{(n-1)})^2}\right) \ .
\label{A7}
\end{equation}
Employing the definition~(\ref{Sol3}) of $|\xi_q^{(n)}|^2$, we arrive
at the final result for $|w_{p}^{(n)}|^2$ in
Eq.~(\ref{Sol2}). Equation~(\ref{rec12}) and the corresponding
equation for the odd levels are evaluated similarly, yielding the
results stated in Sec.~\ref{sec24}. Equation (\ref{rec13}) has to be
treated differently due to the Grassmann singularities, occurring on
the left hand side.  Inserting the expressions Eq.~(\ref{Sol1}) into
Eq.~(\ref{rec13}) we have
\begin{eqnarray}
1 &=& |\beta_p^{(n)}|^2 \left(\sum_m^{\frac{k_1-n+1}{2}}
           \frac{(s_{m1}^{(n-1)})^2+(is_{p2}^{(n)})^2}
   {\left((s_{m1}^{(n-1)})^2-(is_{p2}^{(n)})^2\right)^2}|v_m^{(n)}|^2
          +\sum_{\stackrel{m^\prime=1}{m^\prime\neq p} }^{k_2}
          \frac{(is_{m^\prime2}^{(n-1)})^2+(is_{p2}^{(n)})^2}
         {\left((is_{m^\prime2}^{(n-1)})^2-(is_{p2}^{(n)})^2\right)^2}
             |\alpha_{m^\prime}^{(n)}|^2 \right.\nonumber\\
 & &    \quad \left.+\ \frac{\prod_{q=1}^{\frac{k_1-n-1}{2}}
                \left((is_{p2}^{(n-1)})^2-(s_{q1}^{(n)})^2\right)
                \prod_{q=1,q\neq p}^{k_2}
                \left((is_{p2}^{(n-1)})^2-(is_{q2}^{(n-1)})^2)\right)}
               {\prod_{q=1,q\neq p}^{k_2}
                \left((is_{p2}^{(n-1)})^2-(is_{q2}^{(n)})^2\right)
                \prod_{q=1}^{\frac{k_1-n+1}{2}}
 \left((is_{p2}^{(n-1)})^2-(s_{q1}^{(n-1)})^2\right)}
                \frac{(is_{p2}^{(n-1)})^2+(is_{p2}^{(n)})^2}
               {{|\xi_p}^{(n)}|^2} \right) \ .
\label{A9}
\end{eqnarray}
To cancel the singularity, $|\beta_p^{(n)}|^2$ has to be expanded in
terms of $c_p^{(n)}|\xi_p^{(n)}|^2 $. The expansion coefficient
$c_p^{(n)}$ now contains a nonzero part and its inverse is therefore
well defined. Dividing both sides by $c_p^{(n)}$ and ordering the
right hand side by powers of $|\xi_p^{(n)}|^2$, one finds
\begin{eqnarray}
\frac{1}{c_p^{(n)}} &=& 2 (is_{p2}^{(n-1)})^2
                \frac{
                \prod_{q=1}^{\frac{k_1-n-1}{2}}
                \left((is_{p2}^{(n-1)})^2-(s_{q1}^{(n)})^2\right)
                \prod_{q=1,q\neq p}^{k_2}
                \left((is_{p2}^{(n-1)})^2-(is_{q2}^{(n-1)})^2)\right)}
               {\prod_{q=1,q\neq p}^{k_2}
                \left((is_{p2}^{(n-1)})^2-(is_{q2}^{(n)})^2\right)
                \prod_{q=1}^{\frac{k_1-n+1}{2}}
                \left((is_{p2}^{(n-1)})^2-(s_{q1}^{(n-1)})^2\right)}+\cr
 & &\quad \left( \sum_m^{\frac{k_1-n+1}{2}}
          \frac{(s_{m1}^{(n-1)})^2+(is_{p2}^{(n)})^2}
          {\left((s_{m1}^{(n-1)})^2-(is_{p2}^{(n)})^2\right)^2}
          |v_m^{(n)}|^2+
          \sum_{\stackrel{m^\prime=1}{m^\prime\neq p} }^{k_2}
          \frac{(is_{m^\prime2}^{(n-1)})^2+(is_{p2}^{(n)})^2}
          {\left((is_{m^\prime2}^{(n-1)})^2-(is_{p2}^{(n)})^2\right)^2}
             |\alpha_{m^\prime}^{(n)}|^2 \right.\cr
 & &\quad \left. -\frac{\prod_{q=1}^{\frac{k_1-n-1}{2}}
                \left((is_{p2}^{(n-1)})^2-(s_{q1}^{(n)})^2\right)
                \prod_{q=1,q\neq p}^{k_2}
                \left((is_{p2}^{(n-1)})^2-(is_{q2}^{(n-1)})^2)\right)}
               {\prod_{q=1,q\neq p}^{k_2}
                \left((is_{p2}^{(n-1)})^2-(is_{q2}^{(n)})^2\right)
                \prod_{q=1}^{\frac{k_1-n+1}{2}}
         \left((is_{p2}^{(n-1)})^2-(s_{q1}^{(n-1)})^2\right)}\right)
                |\xi_p^{(n)}|^2 \ .
\label{A10}
\end{eqnarray}
Since $c_p^{(n)}$ and thus $1/c_p^{(n)}$ are of order zero in
$|\xi_p^{(n)}|^2$, the whole term in round brackets can be
neglected. It can be shown by straightforward manipulations that this
term leads just to a shift of $(is_{p2}^{(n-1)})^2 \rightarrow
(is_{p2}^{(n)})^2$ in the resulting expression for $c_p^{(n)}$. This
does not affect $|\beta_p^{(n)}|^2$. Hence, we immediately arrive at
the result for $|\beta_p^{(n)}|^2$ given in Eq.~(\ref{Sol2}).  The
equations for the odd levels are treated accordingly.

\section{Real form of the projection matrices $\widehat{b}^{(n)}$}
\label{appB}

We restrict ourselves to the case $n\leq k_1$, $(k_1-n+1)$ even. The
odd case can be treated accordingly.  The rectangular
$(k_1-n+1+k_2)\times(k_1-n+k_2)$ matrix $\widehat{b}^{(n)T}$ can
schematically be written as
\begin{equation}
\widehat{b}^{(n)}=\left[
  \matrix{\tilde{b}_{11}^{(n)}&\tilde{b}_1^{(n)}&\tilde{b}_{12}^{(n)}\cr
          \tilde{b}_{21}^{(n)}&\tilde{b}_2^{(n)}&\tilde{b}_{22}^{(n)}
         }
         \right]\ .
\label{B1}
\end{equation}
Here, $\tilde{b}_{11}^{(n)}$ is a $(k_1-n+1)/2 \times (k_1-n-1)/2$
matrix with entries
\begin{equation}
(\tilde{b}_{11}^{(n)})_{ij}=\sqrt{2}\frac{|v_i^{(n)}||w_j^{(n)}|}
                          {(s_{i1}^{(n-1)})^2-(s_{j1}^{(n)})^2}
                            \left[
                            \matrix{s_{j1}^{(n)} \cos \vartheta_i^{(n)}&
                            s_{j1}^{(n-1)} \sin \vartheta_i^{(n)}\cr
                            -s_{j1}^{(n)} \sin \vartheta_i^{(n)}&
                            s_{j1}^{(n-1)} \cos \vartheta_i^{(n)}}
                            \right]\quad .
\label{B2}
\end{equation}
The matrix $\tilde{b}_{12}^{(n)}$ has dimension $(k_1-n+1)/2 \times
k_2$ and entries
\begin{eqnarray}
(\tilde{b}_{12}^{(n)})_{ij}&=&
           \frac{|v_i^{(n)}|}{(s_{i1}^{(n-1)})^2-(is_{j2}^{(n)})^2}
                         \nonumber\\
      & & \qquad \left[
  \matrix{\beta_j^{(n)*}\left(is_{j2}^{(n)} \cos \vartheta_i^{(n)}+
            s_{j1}^{(n-1)}i \sin\vartheta_i^{(n)}\right)&
           -i\beta_j^{(n)*} \left(s_{i1}^{(n-1)} \cos \vartheta_i^{(n)}+
            is_{j2}^{(n)} i \sin \vartheta_i^{(n)}\right)\cr
           -\beta_j^{(n)}\left(is_{j2}^{(n)} \cos \vartheta_i^{(n)}-
           s_{j1}^{(n-1)}i \sin \vartheta_i^{(n)}\right)&
           -i\beta_j^{(n)} \left(s_{i1}^{(n-1)}\cos\vartheta_i^{(n)}-
                       is_{j2}^{(n)} i \sin \vartheta_i^{(n)}\right)}
           \right] \ .
\label{B3}
\end{eqnarray}
Moreover, $\tilde{b}_{21}^{(n)}$ is a $k_2 \times (k_1-n+1)/2$ matrix
with entries
\begin{equation}
(\tilde{b}_{21}^{(n)})_{ij}=\frac{|w_j^{(n)}|}
                                 {(is_{i2}^{(n-1)})^2-(s_{j1}^{(n)})^2}
                      \left[
                      \matrix{\alpha_i^{(n)} s_{j1}^{(n)} &
                             i\alpha_i^{(n)} is_{i2}^{(n-1)}\cr
                              \alpha_i^{(n)*}s_{j1}^{(n)} &
                            -i\alpha_i^{(n)*} is_{i2}^{(n-1)}}
                      \right] \ ,
\label{B4}
\end{equation}
and $\tilde{b}_{22}^{(n)}$ is a $k_2 \times (k_1-n+1)/2$ matrix
with entries
\begin{equation}
(\tilde{b}_{22}^{(n)})_{ij}=\sqrt{2}
                      \left[
        \matrix{\alpha_i^{(n)}\beta_j^{(n)*}/
                (is_{i2}^{(n-1)}-is_{j2}^{(n)}) &
                \alpha_i^{(n)}\beta_j^{(n)}/
                (is_{i2}^{(n-1)}+is_{j2}^{(n)}) \cr
                -\alpha_i^{(n)*}\beta_j^{(n)*}/
                (is_{i2}^{(n-1)}+is_{j2}^{(n)}) &
                -\alpha_i^{(n)*}\beta_j^{(n)}/
                (is_{i2}^{(n-1)}-is_{j2}^{(n)})
                }
                \right] \ .
\label{B5}
\end{equation}
Finally, the entries of $\tilde{b}_1^{(n)}$ and $\tilde{b}_2^{(n)}$ are
given by
\begin{eqnarray}
(\tilde{b}_1^{(n)})_i &=&
            \sqrt{2}\frac{|v_i^{(n)}||w_{\frac{k_1-n+1}{2}}^{(n)}|}
                                   {s_{i1}^{(n-1)}}
                      \left[\matrix{\sin \vartheta_i^{(n)}\cr
                                   \cos \vartheta_i^{(n)}}\right] \ ,
               \quad  i=1,\ldots\frac{k_1-n+1}{2} \ ,
                           \nonumber\\
(\tilde{b}_2^{(n)})_i &=&
             \frac{1}{\sqrt{2}}
                       \frac{|w_{\frac{k_1-n+1}{2}}^{(n)}|}
                            {s_{i1}^{(n-1)}}
                      \left[\matrix{i \alpha_i^{(n)}\cr
                                    -i\alpha_i^{(n)*}}\right] \ ,
               \quad i=1,\ldots,k_2 \ .
\label{B6}
\end{eqnarray}
We notice that all elements of $\widehat{b}^{(n)}$ are real.


\begin{thebibliography}{30}

\bibitem{GT}          I.M. Gelfand and M.L. Tzetlin,
                      Dokl. Akad. Nauk. {\bf 71}, 825 (1950)
\bibitem{GT2}         I.M. Gelfand and M.L. Tzetlin,
                      Dokl. Akad. Nauk. {\bf 71}, 1017 (1950)
\bibitem{BR}          A.O. Barut and R. Raczka,
                      Theory of Group Representations and Applications,
                      Warszawa: Polish Scientific Publishers, 1980
\bibitem{SLS}         S.L. Shatashvili,
                      Commun. Math. Phys. {\bf 154}, 421 (1993)
\bibitem{GGT}         T. Guhr,
                      Commun. Math. Phys. {\bf 176}, 555 (1996)
\bibitem{GS1}         V. Guillemein and S. Sternberg,
                      J. Functional Analysis. {\bf 52}, 106 (1983)
\bibitem{GS2}         V. Guillemein and S. Sternberg,
                      Symplectic Techniques in Physics,
                      Cambridge: University Press, 1984
\bibitem{AFS}         A. Alekseev, L. Fadeev and S.L. Shatashvili,
                      J. Geom. Phys. {\bf 5}, 391 (1989)
\bibitem{GUKOP1}      T. Guhr and H. Kohler,
                      {\tt math-ph/0011007}
\bibitem{GUKO1}       T. Guhr and H. Kohler,
                      J. Math. Phys. {\bf 43}, 2707 (2002)
\bibitem{GUKOP2}      T. Guhr and H. Kohler,
                      {\tt math-ph/0012047}
\bibitem{GUKO2}       T. Guhr and H. Kohler,
                      J. Math. Phys. {\bf 43}, 2741 (2002)
\bibitem{KAC1}        V.C. Kac,
                      Comm. Math. Phys. {\bf 53}, 31 (1977)
\bibitem{KAC2}        V.C. Kac,
                      Advances in Math. {\bf 26}, 8 (1977)
\bibitem{RIT}         V. Rittenberg,
                      A Guide to Lie Superalgebras,
                      Lecture Notes in Physics {\bf 79},
                      Berlin: Springer--Verlag, 1977
\bibitem{BER}         F.A. Berezin,
                      Introduction to Superanalysis,
                      MPAM{\bf 9},
                      Dordrecht: D. Reidel Publishing, 1987
\bibitem{MEHTA2}      M.L. Mehta,
                      Matrix Theory,
                      Les Ulis: Les Editions de Physique, 1989
\bibitem{TG}          T. Guhr,
                      J. Math. Phys. {\bf 32}, 336 (1991)
\bibitem{GUH4}        T. Guhr,
                      Ann. Phys. (NY) {\bf 250}, 145 (1996)
\bibitem{DS}          N. Dunford and J.T. Schwartz,
                      Linear Operators 2 -- Spectral Theory,
                      New York, Wiley, 1988
\bibitem{GN}          I.M. Gelfand and M.A. Naimark,
                      Trudi MIAN {\bf 36}, 1 (1950)
\bibitem{GUH5}        T. Guhr,
                      J. Math. Phys. {\bf 34}, 2541 (1993)

\end{thebibliography}
\end{document}